\documentclass[showpacs,preprintnumbers,aps,prl,letterpaper,superscriptaddress,nofootinbib,tightenlines,floats,floatfix]{revtex4-2}
\usepackage{graphicx}
\usepackage{bm}
\usepackage{latexsym}
\usepackage{amsmath,amssymb}
\usepackage{amsfonts}
\usepackage[draft=false]{hyperref}
\usepackage[latin1]{inputenc}
\usepackage{amsthm}
\usepackage{mathrsfs}
\usepackage{ifpdf}
\usepackage{color}
\usepackage{epstopdf}
\usepackage[skip=2pt,font=scriptsize]{caption}
\usepackage{natbib}
\usepackage{blindtext}
\usepackage{tikz}
\usepackage{braket}
\DeclareMathOperator{\Tr}{Tr}
\allowdisplaybreaks

\begin{document}
\title{Effective spherical symmetry in Loop Quantum Gravity: A path integral approach} 

\author{Juan Carlos Del \'Aguila}
\email{jcdelaguila@xanum.uam.mx}
\affiliation{Departamento de F\'isica, Universidad Aut\'onoma Metropolitana Iztapalapa, San Rafael Atlixco 186, CP 09340, Ciudad de M\'exico,
  M\'exico.}
\author{Hugo A. Morales-T\'ecotl}
\email{hugo@xanum.uam.mx}
\affiliation{Departamento de F\'isica, Universidad Aut\'onoma Metropolitana Iztapalapa, San Rafael Atlixco 186, CP 09340, Ciudad de M\'exico,
  M\'exico.}

\begin{abstract}
In this work a loop quantum corrected model is obtained for spherically symmetric space-times in vacuum. This effective model is derived by the use of the path integral method, previously employed in several models of Loop Quantum Cosmology. Our principal aim is to find explicit corrections corresponding to inverse triad and holonomy effects that commonly arise from the loop quantization procedure. These corrections modify the Hamiltonian constraint of the classical theory, adding quantum parameters that represent the length of the holonomies considered during quantization. The semiclassical theory yielded reduces to the classical case when small values of such length are taken to be small. Solutions to the effective dynamics of a simplified version of the complete corrected theory are then found and used to describe an effective geometry with inverse triad corrections. This modified space-time represents a black hole with a curvature singularity in its interior which, contrary to its classical counterpart, does not lead to null geodesic incompleteness. For the case of holonomy corrections, preliminary arguments are given in favor of a potential singularity resolution.
\end{abstract}

\date{Received: date / Accepted: date}

\maketitle

\date{\today}


\maketitle

\section{I. Introduction}

The quantization of gravity is one of the most important questions nowadays in theoretical Physics. While its classical character is majorly well-understood as described by General Relativity (GR), its quantum theory remains to be completely well established at a theoretical level. Loop Quantum Gravity (LQG) has risen as one of the leading proposals that attempt to quantize gravity. At the core of the theory is the discreteness of geometry, which leads to important implications whose physical predictions differ from other canonical quantizations such as Wheeler-DeWitt theory (see for example \cite{WDW, Kiefer} and references therein). The most relevant of the mentioned predictions being the replacement of classical singularities by a big bounce in cosmological models. Indeed, Loop Quantum Cosmology (LQC) has yielded results suggesting singularity avoidance in spatially homogeneous space-times, be them isotropic \cite{LQC, Improved} or anisotropic \cite{SchwarzschildSingularity, Modesto, Revisited}. The standard procedure in those analysis involves working from the onset in the so-called mini-superspace, where only a finite number of gravitational degrees of freedom remains. 

Due to the complications that arise during the quantum description, even within the simplest mini-superspaces, alternative approaches that allow a direct comparison with a well-known theory such as GR have been proposed. These are the so-called effective models in LQG, which introduce corrections that represent loop quantum effects into the classical setting. They can be considered as modified theories of gravity, leading to a semiclassical approximation of the complete quantization. Some of the techniques that have been implemented to obtain effective models are related to the computation of expectation values of operators that have relevant classical analogues. For example, the Hamiltonian constraint operator using Gaussian states \cite{LTB}, and geometric operators associated to the metric components \cite{Gambini2020}. Another interesting scheme is that of the path integral approach, which will be the main purpose of this paper. These effective methods have been extensively applied to LQC models and so far, the results found are in agreement with the big bounce conclusion previously obtained by studying the properties of the Hamiltonian operator. Some interesting results of such effective treatments are the boundedness of scalar quantities relevant to the appearance of singularities such as curvature invariants, expansion, and shear in homogeneous cosmologies \cite{Corichi, Joe, Hugo2}, and the resolution of those singularities as a big bounce \cite{Improved, RobustnessBounce}. Regarding black holes, similar conclusions have been derived with effective techniques, for instance, \cite{AOS,AOS2}.

The path integral method was first considered within the context of LQC in the seminal reference \cite{PathIntegralSLQC} for the homogeneous and isotropic model. Since then, it has also seen use for anisotropic space-times such as Kantowski-Sachs \cite{Hugo1} and Bianchi I \cite{BianchiI}. The main idea consists in writing the transition amplitude between quantum geometry states in the Hamiltonian theory as an integral over all possible classical configurations weighted by the exponential of the action. The loop quantized Hamiltonian constraint operator is used in the process, which guarantees that the underlying discrete geometry of the theory is being captured. Despite the fact that the starting point of the calculation is a discrete orthonormal basis of kinematic states describing the quantum geometry, the path integral can be reformulated in terms of continuous phase space variables taking values over the whole real space. Thereby the argument of the exponential in the weight factor can then be interpreted as an effective action in the framework of a classical theory. By recalling that in a semiclassical saddle point approximation, the amplitude is dominated by such an action, one gets that the ensuing effective equations of motion replace the classical ones. This makes the path integral scheme a well-suited tool in the objective of finding semiclassical models containing loop quantization effects.

On the other hand, there is a natural interest to study spherically symmetric models in LQG due to the various applications they may have, for instance, in the description of Hawking radiation and the exterior regions of black holes. However, the analysis in these models becomes significantly more complex than that those of LQC, mainly due to the former being spatially inhomogeneous. Because of this last property, field aspects of the theory arise that were not part of the homogeneous cosmologies. In fact, spherically symmetric models are sometimes referred to as midi-superspace models \cite{midi}. The first comprehensive treatments of this type of space-times in the context of a canonical formulation aimed at quantization can be found in \cite{Thiemann, Kuchar}. The symmetry reduction utilized for loop quantization was developed in \cite{Kastrup}, while the loop quantization itself was performed in \cite{Bojowald2004,Bojowald2006}. At the effective level, one of the consequences of spatial inhomogeneity is that the diffeomorphism invariance of the model, which now includes a radial part, is not trivial. Such an issue is not troublesome in cosmological settings since only time reparametrizations are allowed within the reduced model. Indeed, previous works report anomalies with the covariance of these inhomogeneous models \cite{Deformed1,Covariance1,Deformed2}. An approach recently taken to overcome this problem has been that of focusing on the covariant aspect of the effective model from the start, and then constructing a spherically symmetric system that has diffeormophism invariance as a defining property. It has turned out that these covariant models describe non-singular black hole metrics \cite{asier,belfaqih,zhang}. 

In this paper we wish to apply the path integral method to spherically symmetric space-times, obtaining thus its corresponding semiclassical model. Our focus will shift then to the covariance of the model, or lack thereof, and whether or not the classical singularity is avoided. The work is structured as follows. In section II, a classical description of the spherically symmetric model is presented, along with results of its loop quantization. Section III is dedicated to explaining the process of applying the path integral scheme to the geometries of interest. Here, the quantum effective model is found, and then its properties and possible issues are commented in section IV, this leads us to consider a simplified version of it. This simplified model, and in particular, its solutions to the effective dynamics in different gauges, are studied in section V. The physical interpretation of those solutions is discussed in section VI, where an effective space-time is constructed and the consequences of quantum effects are analyzed. Conclusions and some discussion are elaborated in the final section of the paper.  

\section{II. The Spherically Symmetric Model}

Let us introduce first the classical description in the Hamiltonian formulation of spherically symmetric space-times. Here only a brief initial summary will be presented, helpful references for a more in-depth discussion of the subject are \cite{Bojowald2004, Campiglia2007}.

\subsection{II.A. Classical Description}

Using adapted Ashtekar-Barbero variables, the phase space of the spherically symmetric model is characterized by two degrees of freedom which we express as, $\mathbf{\Gamma}=(\bar{K}_r,\bar{K}_\varphi ; E^r, E^\varphi)$. In this case, $\bar{K}_r=\gamma K_r$ and $\bar{K}_\varphi=\gamma K_\varphi$ are proportional to the components $K_r$ and $K_\varphi$ of the extrinsic curvature term that appears in the Ashtekar connection, and $\gamma$ is the Barbero-Immirzi parameter. A bar over the past quantities is employed to denote this relationship. The components $E^r$ and $E^\varphi$ of the densitized triad of the spatial metric are respectively the conjugate momenta of the configuration variables $\bar{K}_r$ and $\bar{K}_\varphi$. This information is then complemented by the lapse function $N$, and the only non-trivial component $N_r$ of the shift vector, to fully describe the geometry of the spacetime. In spherical coordinates $\{t,r,\theta,\varphi\}$ the line element can be written as

\begin{equation}
ds^2=-N^2dt^2+\frac{(E^\varphi)^2}{E^r}\left(dr+N_rdt\right)^2+E^rd\Omega^2,
\label{ds2Esf}
\end{equation}
where $d\Omega^2$ is the line element of the two-sphere. Since we wish to describe the most general spherically symmetric space-time, all of the phase space variables depend on the time and radial coordinates, $t$ and $r$. The Hamiltonian is a sum of two constraints\footnote{During the analysis with Ashtekar-Barbero variables, a non-trivial component of the Gauss constraint would appear as an extra term in the total Hamiltonian. In the expressions (\ref{Dc}) and (\ref{Hc}) for the Hamiltonian and diffeormorphism constraints, this Gauss component has already been solved, thus reducing the phase space to that which was previously introduced, $\mathbf{\Gamma}=(\bar{K}_r,\bar{K}_\varphi ; E^r, E^\varphi)$.} $H_T=H_c[N]+C_c[N_r]$, whose explicit expressions are,

\begin{eqnarray}
C_c[N_r]&=&\frac{1}{2\gamma}\int drN_r\left[2E^\varphi\bar{K}'_\varphi-(E^r)'\bar{K}_r\right], \label{Dc} \\
H_c[N]&=&-\int drN\left[\frac{\bar{K}_\varphi}{\gamma^2\sqrt{|E^r|}}\left(\bar{K}_rE^r+\frac{1}{2}\bar{K}_\varphi E^\varphi\right)+\frac{E^\varphi}{2\sqrt{|E^r|}}(1-\Gamma_\varphi^2)+\sqrt{|E^r|}\Gamma'_\varphi\right],
\label{Hc}
\end{eqnarray}
with $\Gamma_\varphi=-E^{r\prime}/2E^\varphi$. These expressions are known as the diffeomorphism and Hamiltonian constraints, respectively. Also, a prime denotes derivation with respect to $r$. In these variables, the symplectic structure is given by the following Poisson brackets,

\begin{equation}
\{\bar{K}_r(t,r),E^r(t,r')\}=2\gamma\delta(r,r'), \quad \{\bar{K}_\varphi(t,r),E^\varphi(t,r')\}=\gamma\delta(r,r').
\label{EstSimp}
\end{equation}
The algebra of constraints that (\ref{Dc}) and (\ref{Hc}) form under such brackets is

\begin{align}
\{C_c[N_r],H_c[N]\}=&H_c[N_rN'], \quad \{C_c[N_r],C_c[M_r]\}=C_c[N_rM'_r-M_rN'_r], \nonumber \\
\{H_c[N],H_c[M]\}=&C_c\left[\frac{E^r}{(E^\varphi)^2}\left(NM'-MN'\right)\right].
\label{CAEsf}
\end{align}

Prior to the loop quantization procedure it is necessary to specify holonomies and fluxes, the elementary variables of LQG, within the classical description of spherically symmetric geometry. These quantities will be given in terms of densitized triads and extrinsic curvature and eventually shall become operators in a suitable Hilbert space. We begin by considering holonomies, i.e., parallel transport operators, along preferred directions in spherical coordinates. In the following, the label $I$ will denote an arbitrary radial interval. The configuration degrees of freedom $\bar{K}_r$ and $\bar{K}_\varphi$ are the components of the connection used to compute them. We thus obtain three different types of holonomies $h_{i}[\bar{K}]$:

\begin{itemize}
    \item Radial with length $\lambda$, 
    \begin{equation}
         h_r[\bar{K}]=e^{I_r\tau_3}=\cos\left(\frac{1}{2}I_r\right)+2\tau_3\sin\left(\frac{1}{2}I_r\right), \quad I_r=\int_I\bar{K}_rdr.
    \label{hr}
    \end{equation}
    \item Polar (along $\theta$) with length $\mu_\theta$, 
    \begin{equation}
        h_\theta[\bar{K}]=e^{\mu_\theta\bar{K}_\varphi\tau_1}=\cos\left(\frac{1}{2}\mu_\theta\bar{K}_\varphi\right)+2\tau_1\sin\left(\frac{1}{2}\mu_\theta\bar{K}_\varphi\right)
    \label{htheta}
    \end{equation}
    \item Azimuthal in the equator (along $\varphi$ with $\theta=\pi/2$) with length $\mu_\varphi$, 
    \begin{equation}
         h_\varphi[\bar{K}]=e^{-\mu_\varphi\bar{K}_\varphi\tau_2}=\cos\left(\frac{1}{2}\mu_\varphi\bar{K}_\varphi\right)-2\tau_2\sin\left(\frac{1}{2}\mu_\varphi\bar{K}_\varphi\right).
    \label{hphi}
    \end{equation}
\end{itemize}
Here $\tau_i=-i\sigma_i/2$ is a basis for the $su(2)$ algebra, and $\sigma_i$ are the Pauli matrices.

In a similar manner, fluxes $E(S_{ij})$ through three preferred surfaces in space will be given by the components of densitized triads $E^r$ and $E^\varphi$. Namely:

\begin{itemize}
    \item Surface of constant $\varphi$, $$E(S_{r\theta})\sim\int_I E^\varphi dr.$$
    \item Surface of constant $r$, $$E(S_{\theta\varphi})\sim E^r.$$
    \item Surface of constant $\theta=\pi/2$, $$E(S_{r\varphi})\sim\int_I E^\varphi dr.$$
\end{itemize}

\subsection{II.B. Loop Quantization}

The main ingredients for the loop quantization of spherically symmetric geometry will be described in this subsection. A detailed treatment of the scheme can be found in \cite{Bojowald2004,Bojowald2006}.

\begin{figure}[h]
\centering
\begin{tikzpicture} \centering
    \filldraw (-0.45,0) circle (0.5pt);
    \filldraw (-0.3,0) circle (0.5pt);
    \filldraw (-0.15,0) circle (0.5pt);
    \draw (0,0) -- (4,0) node[anchor=north west, xshift=-2.95cm, yshift=0.45cm] {\scriptsize $k_{i-1}$} node[anchor=north west, xshift=-1.8cm, yshift=0.45cm] {\scriptsize $k_i$};
    \filldraw (1,0) circle (2pt) node[anchor=north west, xshift=-0.5cm, yshift=0cm] {\scriptsize $\mu_{i-1}$};
    \filldraw (2,0) circle (2pt) node[anchor=north west, xshift=-0.25cm, yshift=0cm] {\scriptsize $\mu_i$};
    \filldraw (3,0) circle (2pt) node[anchor=north west, xshift=-0.35cm, yshift=0cm] {\scriptsize $\mu_{i+1}$};
    \filldraw (4.15,0) circle (0.5pt);
    \filldraw (4.3,0) circle (0.5pt);
    \filldraw (4.45,0) circle (0.5pt);
\end{tikzpicture}
\caption{A graph in the radial direction with labels $\mu$ and $k$ for its vertices and edges, respectively.}
\label{g}
\end{figure}
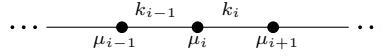

An orthonormal basis $T_{g,k,\mu}$ of the kinematical Hilbert space is given by the so-called spin network states, which can be constructed from the holonomies (\ref{hr}) to (\ref{hphi}), for instance, by combining the preferred directions and therefore forming closed loops. Explicitly,

\begin{equation}
    T_{g,k,\mu}=\prod_{e\in g}\exp{\left[\frac{i}{2}k_e\int_e\bar{K}_rdr\right]}\prod_{v\in V(g)}\exp{\left[\frac{i}{2}\mu_v\bar{K}_\varphi\right]}.
    \label{redspin}
\end{equation}
To express such a spin network state, we are considering an arbitrary graph $g$ in the radial direction. The edges of the graph are represented by $e$ and its vertices by $v$. The set of all vertices in $g$ is written as $V(g)$. The parameter $k_e\in\mathbb{Z}$ is a label that characterizes edges, while $\mu_v\in\mathbb{R}$ characterizes vertices. It is important to highlight that the states (\ref{redspin}) are already gauge invariant under $SU(2)$ transformations generated by the Gauss constraint. The spin network states for a given graph $g$ can be visualized as shown in figure \ref{g}.

In the momentum representation, the state corresponding to the graph shown in figure \ref{g} can then be written as

\begin{equation}
    \ket{\ldots,\mu_{i-1},k_{i-1},\mu_i,k_i,\mu_{i+1},\ldots}=\ket{\vec{\mu},\vec{k}}.
    \label{muk}
\end{equation}
The right hand side of equation (\ref{muk}) will be used as a compact way to write a general state, for which the following holds $$\braket{\bar{K}_r,\bar{K}_\varphi | \vec{\mu},\vec{k}}=T_{g,k,\mu}.$$

The next step is to promote the previously described holonomy and fluxes to operators acting on the kinematical Hilbert space. We will start with the conjugate momenta $E^r$ and $E^\varphi$, which will act multiplicatively and for each vertex in the graph,

\begin{equation}
    \int_I\hat{E}^\varphi dr\ket{\vec{\mu},\vec{k}}=\gamma l_{Pl}^2\sum_{v\in I}\mu_v\ket{\vec{\mu},\vec{k}}, \quad \hat{E}^r(v_i)\ket{\vec{\mu},\vec{k}}=\frac{1}{2}\gamma l_{Pl}^2\left(k_i+k_{i-1}\right)\ket{\vec{\mu},\vec{k}}.
    \label{E}
\end{equation}
Here, the Planck length is given by $l_{Pl}$. It is also possible to express a distributional version of the first flux operator, this is, $$\hat{E}^\varphi\ket{\vec{\mu},\vec{k}}=\gamma l_{Pl}^2\sum_{v\in I}\delta(r,v)\mu_v\ket{\vec{\mu},\vec{k}}.$$ It will also be important to construct a volume operator $\hat{V}$, its classical counterpart can be written as $$V(I)=4\pi\int_I|E^\varphi|\sqrt{|E^r|}dr.$$ The volume $V(I)$ can then be readily promoted to flux operators, yielding $$\hat{V}(I)=4\pi\int_I|\hat{E}^\varphi|\sqrt{|\hat{E}^r|}dr.$$ Hence, using (\ref{E}) we have that,
\begin{equation}
    \hat{V}(I)\ket{\vec{\mu},\vec{k}}=\sum_{v\in I}V_{\vec{\mu},\vec{k}}\ket{\vec{\mu},\vec{k}}, \quad V_{\vec{\mu},\vec{k}}=4\pi\sqrt{\gamma^3}l_{Pl}^3|\mu_i|\sqrt{\frac{1}{2}\left|k_i+k_{i-1}\right|}.
    \label{V}
\end{equation}
It is easy to note that, due to the fact that the flux operators are diagonal for the orthonormal basis $\ket{\vec{\mu},\vec{k}}$, the volume will also share this property. This considerably simplifies the action of the quantized Hamiltonian constraint on such states.

Now the honolomy operators, $h_r(k_i)$ and $h_\varphi(v_i,\delta_\varphi)$, will be introduced. For the sake of lightening notation, the $\bar{K}$ dependence on $h_i[\bar{K}]$ has been dropped and replaced by additional parameters that specify on which vertex or edge the operators act. In particular, the path of the radial holonomy $h_r(k_i)$ is taken along the the edge $k_i$ connecting the $v_i$ and $v_{i+1}$ vertices. Meanwhile, the point-holonomy $h_\varphi(v_i,\delta_\varphi)$ denotes parallel transport in the $v_i$ vertex along the $\varphi$ direction with a curve of length $\delta_\varphi$, this will be referred to as the holonomy parameter. Both of them shift the labels of their corresponding vertices and edges. Respectively, we have that,

\begin{equation}
    \widehat{\exp}{\left[\frac{i}{2}\int_{k_i}\bar{K}_rdr\right]}\ket{\vec{\mu},\vec{k}}=\ket{\vec{\mu},\vec{k}+\vec{k}_i}, \quad \hat{e}^{\frac{i}{2}\delta_\varphi\bar{K}_\varphi(v_i)}\ket{\vec{\mu},\vec{k}}=\ket{\vec{\mu}+\vec{\delta}_i,\vec{k}},
    \label{he}
\end{equation}
where we are using the notation

\begin{equation}
    \ket{\vec{\mu}+\vec{\delta}_i,\vec{k}}=\ket{\ldots,\mu_{i-1},k_{i-1},\mu_i+\delta_\varphi,k_i,\mu_{i+1},\ldots}, \quad \ket{\vec{\mu},\vec{k}+2\vec{k}_i}=\ket{\ldots,\mu_{i-1},k_{i-1},\mu_i,k_i+2,\mu_{i+1},\ldots}.
\end{equation}
Consecutive applications of the holonomy operators defined by (\ref{he}) define super-selection sectors for vertex base values $\mu_i^{(0)}$ and edge base values $k_i^{(0)}$, where $0\leq\mu_i^{(0)}<\delta_\varphi$ and $0\leq k_i^{(0)}<2$. Thereby, the vertex and edge labels can always be written as $\mu_i=\mu_i^{(0)}+n_i\delta_\varphi$ and $k_i=k_i^{(0)}+2m_i$, with $n_i,m_i\in\mathbb{Z}$, such that the $\ket{\vec{\mu},\vec{k}}$ states span a countable basis.

One can verify that the flux and holonomy operators defined above satisfy the holonomy-flux algebra that characterizes the loop representation typical of this type of quantization. Namely,

\begin{equation}
\left[\int_I\hat{E}^\varphi dr,\hat{h}_\varphi(v_i,\delta_\varphi)\right]=\gamma l_{Pl}^2\delta_\varphi\hat{h}_\varphi(v_i,\delta_\varphi), \quad \left[\hat{E}^r,\hat{h}_r(k_i)\right]=\gamma l_{Pl}^2\hat{h}_r(k_i).
\label{HF}
\end{equation}
The algebra in (\ref{HF}) replaces the canonical one since, due to the discreteness of the spatial geometry, there are no infinitesimal generators of translations and consequently, the corresponding $\hat{\bar{K}}_r$ and $\hat{\bar{K}}_\varphi$ operators do not exist within the kinematical Hilbert space.

Once that we know how these basic operators act, we can proceed to examine the quantized Hamiltonian constraint of the model. It is performed by dividing said constraint into two terms $H_c[N]=H_K[N]+H_\Gamma[N]$, where

\begin{align}
    H_K[N]=&-\int dr\frac{N\bar{K}_\varphi}{\gamma^2\sqrt{|E^r|}}\left[\bar{K}_rE^r+\frac{1}{2}\bar{K}_\varphi E^\varphi\right], \nonumber\\
    H_\Gamma[N]=&-\int drN\left[\frac{E^\varphi}{2\sqrt{|E^r|}}(1-\Gamma_\varphi^2)+\sqrt{|E^r|}\Gamma'_\varphi\right], \quad \Gamma_\varphi=-\frac{E^{r\prime}}{2E^\varphi}. \nonumber
    \label{HGamma}
\end{align}
The treatment of the $H_K[N]$ term is performed by considering the typical regularization proposed by Thiemann \cite{Thiemann_2007}, which consists on approximating curvature via the holonomies along closed curves $\alpha$ in the limit when they shrink to a vertex $v(\alpha)$. Furthermore, the inverse factor $1/\sqrt{q}$ is replaced by a classical identity in phase space that involves the Possion bracket between holonomies and the volume $V$. The Poisson brackets are then promoted to commutators between operators as suggested by the Dirac prescription. Using geometrical units in which $\hbar=l_{Pl}^2$, this whole non-trivial process yields, 

\begin{equation}
    \hat{H}_K=\sum_v\frac{i}{2\pi\gamma^3\delta_r\delta_\varphi^2l_{Pl}^2}N(v)\Tr\left[\varepsilon^{ijk}\hat{h}_{ij}\hat{h}_k[\hat{h}_k^{-1},\hat{V}]\right],
		\label{OpHK}
\end{equation}
where the quantities $\delta_k$ are holonomy parameters and their index, as well as that in $h_k$, refers to the holonomy along the $x^k$ direction. Thus, $\delta_1=\delta_r$ is the length of a radial edge and $\delta_2=\delta_3=\delta_\varphi$ is the previously introduced parameter. Also, the notation $h_{ij}=h_ih_jh_i^{-1}h_j^{-1}$ is being used to describe the holonomy along a closed loop $\alpha_{ij}$. 

Since the indices $i,j,k$ are being summed and multiplied by the Levi-Civita symbol in equation (\ref{OpHK}), there are three different types of loops. Each of them contribute twice to the sum, the first one when its path is traversed along a certain direction and the second one in the reverse case. This gives a total of six terms in which the parameter $\delta_i$ is always the same for all holonomies $h_i$, independently of the loop considered. In the following our analysis will slightly differ from this homogeneous property and shall consider the possibility of different holonomy parameters for different types of loops. Namely, we choose the angular path of $\hat{\bar{h}}_{\theta\varphi}$ such that, 

\begin{equation}
\hat{\bar{h}}_{\theta\varphi}=\hat{h}_\theta\left(v_i,\frac{\delta_\varphi}{2}\right)\hat{h}_\varphi\left(v_i,\frac{\delta_\varphi}{2}\right)\hat{h}_\theta^{-1}\left(v_i,\frac{\delta_\varphi}{2}\right)\hat{h}_\varphi^{-1}\left(v_i,\frac{\delta_\varphi}{2}\right),
\label{htpmod}
\end{equation}
while the other closed holonomies remain as $$\hat{h}_{r\theta}=\hat{h}_r(k_i)\hat{h}_\theta(v_{i+1},\delta_\varphi)\hat{h}_r^{-1}(k_i)\hat{h}_\theta^{-1}(v_i,\delta_\varphi), \quad \hat{h}_{\varphi r}=\hat{h}_\varphi(v_i,\delta_\varphi)\hat{h}_r(k_i)\hat{h}_\varphi^{-1}(v_{i+1},\delta_\varphi)\hat{h}_r^{-1}(k_i).$$ The bar over $\hat{\bar{h}}_{\theta\varphi}$ is used to denote that for this loop the holonomy parameter $\delta_\varphi$ is being halved. The reason behind this seemingly unnecessary modification will be explained once we obtain the effective model yielded by this quantization scheme (see subsection IV.A). Equation (\ref{OpHK}) becomes thus,

\begin{equation}
\hat{H}_K=\sum_v\frac{i}{2\pi\gamma^3\delta_r\delta_\varphi^2l_{Pl}^2}N(v)\Tr\left[(\hat{h}_{r\theta}-\hat{h}_{\theta r})\hat{h}_\varphi[\hat{h}_\varphi^{-1},\hat{V}]+(\hat{h}_{\varphi r}-\hat{h}_{r\varphi})\hat{h}_\theta[\hat{h}_\theta^{-1},\hat{V}]+4(\hat{\bar{h}}_{\theta\varphi}-\hat{\bar{h}}_{\varphi\theta})\hat{h}_r[\hat{h}_r^{-1},\hat{V}]\right],
\label{HKmod}
\end{equation}
where the distinct numerical factor of the last term is due to the scaling in the size of the angular loop $\alpha_{\theta\varphi}$, this replaces the factor $1/\delta_\varphi^2$ with $4/\delta_\varphi^2$. The parameters of the individual holonomy operators that appear inside commutators with the volume are left unaltered with respect to the usual case (unscaled parameters). Finally, it is straightforward to realize that the holonomy operator corresponding to $\hat{h}_\theta(v_i,\delta_\varphi/2)$ is $\hat{e}^{\frac{i}{4}\delta_\varphi\bar{K}_\varphi(v_i)}$, which applied to the $\ket{\vec{\mu},\vec{k}}$ states yields, $$\hat{e}^{\frac{i}{4}\delta_\varphi\bar{K}_\varphi(v_i)}\ket{\vec{\mu},\vec{k}}=\Ket{\vec{\mu}+\frac{1}{2}\vec{\delta}_i,\vec{k}}.$$ This consequently changes the superselection sector to $\mu_i=\mu_i^{(0)}+n_i\delta_\varphi/2$, without further implications.  

On the other hand, the second term $H_\Gamma[N]$ can be quantized by noting that it only depends on conjugate momenta, which have well-defined operators in the kinematical Hilbert space. They can be directly promoted to their quantum operators counterparts, except for the root $\sqrt{|E^r|}$ and its inverse $1/\sqrt{|E^r|}$. Their corresponding operators can be found by using the previously outlined regularization procedure. We obtain thus,

\begin{equation}
\hat{H}_\Gamma=-\sum_v\left[\widehat{\frac{E^\varphi}{2\sqrt{E^r}}}\left(1-\hat{\Gamma}_\varphi^2\right)+\widehat{\sqrt{E^r}}\hat{\Gamma}_\varphi'\right],
\end{equation}
with $$\widehat{\frac{E^\varphi}{\sqrt{E^r}}}=-\frac{i}{2\pi\gamma\delta_rl_{Pl}^2}\Tr\left[\hat{h}_r[\hat{h}_r^{-1},\hat{V}]\tau_3\right], \quad \widehat{\sqrt{E^r}}=\frac{i}{4\pi\gamma\delta_\varphi l_{Pl}^2}\Tr\left[\hat{h}_\varphi[\hat{h}_\varphi^{-1},\hat{V}]\tau_2\right].$$ The quantities related to $\Gamma_\varphi$ cannot be expressed only through Poisson brackets, hence, its operators are not immediately obtained by this method. Since they contain radial derivatives, they can be discretized by considering differences between neighboring vertices. An explicit version of $\hat{\Gamma}_\varphi$ will depend on the discretization utilized. Regardless, this operator is always diagonal in the momentum basis used so far, and its eigenvalues $\Gamma_\varphi(\vec{\mu},\vec{k})$ will be in terms of the labels of both vertices and edges involved in the discretization. In what follows, a basic proposal of such a procedure is introduced and its explicit eigenvalues are found.

Since $\Gamma_\varphi=E^{r\prime}/2E^\varphi$, we need to find a suitable operator for $(E^\varphi)^{-1}$. This can be done by again considering identities in the classical phase space that involve Poisson brackets and promoting them to operators. For instance, $$(E^\varphi)^{n-1}=\frac{2}{n\gamma\delta_\varphi}\Tr\left[h_\varphi\{h_\varphi^{-1},(E^\varphi)^n\}\tau_2\right]$$ yields the operator $$(\hat{E}^\varphi)^{n-1}=-\frac{2i}{n\delta_\varphi\gamma l_{Pl}^2}\Tr\left[\hat{h}_\varphi[\hat{h}_\varphi^{-1},(\hat{E}^\varphi)^n]\tau_2\right],$$ for real, non-zero $n$. Using $(\hat{E}^\varphi)^{n-1}$ as defined in the past equation, it can be shown that $$(\hat{E}^\varphi)^{n-1}\ket{\vec{\mu},\vec{k}}=\frac{(\gamma l_{Pl}^2)^{n-1}}{n\delta_\varphi}\left(|\mu+\delta_\varphi|^n-|\mu-\delta_\varphi|^n\right)\ket{\vec{\mu},\vec{k}}.$$ Thus, by setting $n=1/2$ and making $(\hat{E}^\varphi)^{-1}=\left[(\hat{E}^\varphi)^{-1/2}\right]^2$, a densely-defined operator associated to $(\hat{E}^\varphi)^{-1}$ can be expressed. Its eigenvalues are,

\begin{equation}
(\hat{E}^\varphi)^{-1}\ket{\vec{\mu},\vec{k}}=\frac{4}{\delta_\varphi^2\gamma l_{Pl}^2}\left(\sqrt{|\mu+\delta_\varphi|}-\sqrt{|\mu-\delta_\varphi|}\right)^2\ket{\vec{\mu},\vec{k}}, \nonumber
\end{equation}
which are well-defined for all $\mu\in\mathbb{R}$. For the case of the derivative $E^{r\prime}$, we have that 

\begin{equation}
\hat{E}^{r\prime}(v_i)\ket{\vec{\mu},\vec{k}}=\frac{1}{2}\gamma l_{Pl}^2\left(k_i-k_{i-1}\right)\ket{\vec{\mu},\vec{k}}. \nonumber
\end{equation}
Putting together the past operators, i.e., $\hat{\Gamma}_\varphi=-\hat{E}^{r\prime}(\hat{E}^\varphi)^{-1}/4$, we obtain

\begin{equation}
\hat{\Gamma}_\varphi(v_i)\ket{\vec{\mu},\vec{k}}=\Gamma_\varphi(\mu_i,k_i)\ket{\vec{\mu},\vec{k}}, \quad \Gamma_\varphi(\mu_i,k_i)=-\frac{1}{2\delta_\varphi^2}\left(\sqrt{|\mu_i+\delta_\varphi|}-\sqrt{|\mu_i-\delta_\varphi|}\right)^2\left(k_i-k_{i-1}\right).
\label{GammaOp}
\end{equation}
Similarly, a simple discretization is then used to define the last operator we need, $\hat{\Gamma}'_\varphi(v_i)=\hat{\Gamma}_\varphi(v_i)-\hat{\Gamma}_\varphi(v_{i-1})$, with eigenvalues

\begin{equation}
\hat{\Gamma}_\varphi'(v_i)\ket{\vec{\mu},\vec{k}}=\Gamma_\varphi'(\mu_i,k_i)\ket{\vec{\mu},\vec{k}}, \quad \Gamma_\varphi'(\mu_i,k_i)=\Gamma_\varphi(\mu_i,k_i)-\Gamma_\varphi(\mu_{i-1},k_{i-1}).
\end{equation}
Note that $\Gamma_\varphi'(\mu_i,k_i)$ will explicitly depend not only on edge labels $k_i$, but also on $k_{i-1}$ and $k_{i-2}$ labels as well. This reflects the fact that in the classical theory, $\Gamma'_\varphi$ contains second order radial derivatives of $E^r$. 

We now have all the necessary elements to write the explicit expressions generated by the term $\hat{H}_\Gamma$ of the Hamiltonian constraint. The computation of the action of the operator $\hat{H}_c=\hat{H}_K+\hat{H}_\Gamma$ on orthonormal states is straightforward, though somewhat tedious. The final result is 

\begin{equation}
\hat{H}_c\ket{\vec{\mu},\vec{k}}=\frac{1}{4\pi\gamma l_{Pl}^2}\sum_iN(v_i)\left(\frac{1}{\gamma^2\delta_r\delta_\varphi^2}\hat{H}_K^{(i)}+\hat{H}_\Gamma^{(i)}\right)\ket{\vec{\mu},\vec{k}},
\label{Hcuk}
\end{equation} 
with

\begin{align}
    \hat{H}_K^{(i)}\ket{\vec{\mu},\vec{k}}=\frac{1}{4}\left(\frac{}{}\right.&\left.V_{\vec{\mu}+\vec{\delta}_i,\vec{k}}-V_{\vec{\mu}-\vec{\delta}_i,\vec{k}}\right) \nonumber\\
    \times\left(\frac{}{}\right.&\ket{\vec{\mu}+\vec{\delta}_i-\vec{\delta}_{i+1},\vec{k}-2\vec{k}_i}-\ket{\vec{\mu}+\vec{\delta}_i-\vec{\delta}_{i+1},\vec{k}+2\vec{k}_i} \nonumber\\
    &+\ket{\vec{\mu}-\vec{\delta}_i-\vec{\delta}_{i+1},\vec{k}-2\vec{k}_i}-\ket{\vec{\mu}-\vec{\delta}_i-\vec{\delta}_{i+1},\vec{k}+2\vec{k}_i} \nonumber\\
    &-\ket{\vec{\mu}+\vec{\delta}_i+\vec{\delta}_{i+1},\vec{k}-2\vec{k}_i}+\ket{\vec{\mu}+\vec{\delta}_i+\vec{\delta}_{i+1},\vec{k}+2\vec{k}_i} \nonumber\\
    &-\left.\ket{\vec{\mu}-\vec{\delta}_i+\vec{\delta}_{i+1},\vec{k}-2\vec{k}_i}+\ket{\vec{\mu}-\vec{\delta}_i+\vec{\delta}_{i+1},\vec{k}+2\vec{k}_i}\right) \nonumber\\
    +\frac{1}{2}\left(\frac{}{}\right.&\left.V_{\vec{\mu},\vec{k}+2\vec{k}_i}-V_{\vec{\mu},\vec{k}-2\vec{k}_i}\right)\left(\ket{\vec{\mu}-2\vec{\delta}_i,\vec{k}}-2\ket{\vec{\mu},\vec{k}}+\ket{\vec{\mu}+2\vec{\delta}_i,\vec{k}}\right).
    \label{HK}
\end{align}
The sum over the index $i$ means the sum over all vertices $v$ of the graph. On the other hand, since $\hat{H}_\Gamma^{(i)}$ is expressed in terms of flux operators exclusively, it will be diagonal in the momentum basis. Therefore we can write, $$\hat{H}_\Gamma^{(i)}\ket{\vec{\mu},\vec{k}}=H_\Gamma^{(i)}(\vec{\mu},\vec{k})\ket{\vec{\mu},\vec{k}},$$ where

\begin{equation}
\hat{H}_\Gamma^{(i)}(\vec{\mu},\vec{k})=-\frac{1}{2}\left[\frac{1}{\delta_r}\left(V_{\vec{\mu},\vec{k}+2\vec{k}_i}-V_{\vec{\mu},\vec{k}-2\vec{k}_i}\right)\left(1-\Gamma_\varphi^2(\mu_i,k_i)\right)+\frac{1}{\delta_\varphi}\left(V_{\vec{\mu}+\vec{\delta}_i,\vec{k}}-V_{\vec{\mu}-\vec{\delta}_i,\vec{k}}\right)\Gamma_\varphi'(\mu_i,k_i)\right].
    \label{HGammaO}
\end{equation}
Physical states $\ket{\psi_{phys}}$ are then defined to be those who annihilate the Hamiltonian constraint, i.e., $$\ket{\psi_{phys}}=\sum_{\vec{\mu},\vec{k}}\psi_{\vec{\mu},\vec{k}}\ket{\vec{\mu},\vec{k}}, \quad \hat{H}_c\ket{\psi_{phys}}=0,$$ where the sum is to be understood over the countable basis of the super selection sectors previously defined, i.e., $\mu_i=\mu_i^{(0)}+n_i\delta_\varphi$ and $k_i=k_i^{(0)}+2m_i$ with $n_i,m_i\in\mathbb{Z}$.  This yields $n$ coupled difference equations for the coefficients $\psi_{\vec{\mu},\vec{k}}$, being $n$ the total number of vertices. Such system of equations is extremely difficult, if not impossible (at least by analytic tools), to solve. Thus, instead of searching for solutions representing physical states, we will be interested in using (\ref{Hcuk}) to find a loop quantum effective model through the path integral scheme.

Finally it is important to remark that, as opposed to the described Hamiltonian constraint, a similar construction of an operator corresponding to the diffeomorphism constraint is not generally carried out since solutions to it can be found by group averaging kinematical states, see \cite{AshtekarLewandowski} for more details.

\section{III. The Path Integral Approach}

We follow the seminal work from reference \cite{PathIntegralSLQC} and begin by considering the transition amplitude $A(\vec{\mu}_f,\vec{k}_f;\vec{\mu}_i,\vec{k}_i)$ from an initial state characterized by $\vec{\mu}_i$ and $\vec{k}_i$, to a final one with labels $\vec{\mu}_f$ and $\vec{k}_f$. This amplitude can be found by the group averaging procedure which computes the overall action of the Hamiltonian constraint over the initial state, providing physical states $\ket{\psi_{phys}}$ as a result,

\begin{equation}
A(\vec{\mu}_f,\vec{k}_f;\vec{\mu}_i,\vec{k}_i)=\int d\alpha\braket{\vec{\mu}_f,\vec{k}_f | e^{\frac{i}{\hbar}\alpha\hat{H}_c} | \vec{\mu}_i,\vec{k}_i}.
\label{Afi}
\end{equation}

The standard path integral analysis is then followed, which we hereby summarize. First, the evolution of the quantum states is separated into $\mathcal{N}$ partitions, such that, $$e^{\frac{i}{\hbar}\alpha\hat{H}_c}=\prod_{j=1}^\mathcal{N}e^{\frac{i}{\hbar}\varepsilon\alpha\hat{H}_c}, \quad \varepsilon=\frac{1}{\mathcal{N}}.$$ The $\ket{\vec{\mu},\vec{k}}$ states constitute a complete basis of the kinematical Hilbert space in the super selection sectors, and hence, $$\mathbb{I}=\sum_{\vec{\mu},\vec{k}}\ket{\vec{\mu},\vec{k}}\bra{\vec{\mu},\vec{k}}.$$ This completeness relation is inserted between every partition, yielding $$A(\vec{\mu}_f,\vec{k}_f;\vec{\mu}_i,\vec{k}_i)=\sum_{\vec{\mu}_1,\vec{k}_1,\ldots,\vec{\mu}_{\mathcal{N}-1},\vec{k}_{\mathcal{N}-1}}\int d\alpha\braket{\vec{\mu}_\mathcal{N},\vec{k}_\mathcal{N}| e^{\frac{i}{\hbar}\varepsilon\alpha\hat{H}_c} | \vec{\mu}_{\mathcal{N}-1},\vec{k}_{\mathcal{N}-1}}\ldots\braket{\vec{\mu}_1,\vec{k}_1 | e^{\frac{i}{\hbar}\varepsilon\alpha\hat{H}_c} | \vec{\mu}_0,\vec{k}_0},$$ with $\ket{\vec{\mu}_\mathcal{N},\vec{k}_\mathcal{N}}=\ket{\vec{\mu}_f,\vec{k}_f}$ and $\ket{\vec{\mu}_0,\vec{k}_0}=\ket{\vec{\mu}_i,\vec{k}_i}$. The main objective of this analysis is to obtain a first order model, therefore, only terms up to linear order in $\varepsilon$ are considered. The exponential of the Hamiltonian operator is then approximated as, $$e^{\frac{i}{\hbar}\varepsilon\alpha\hat{H}_c}=\mathbb{I}+\frac{i}{\hbar}\varepsilon\alpha\hat{H}_c+\mathcal{O}(\varepsilon^2).$$ 

It is at this point where it becomes essential to have previously computed the action of the Hamiltonian constraint on the $\ket{\vec{\mu},\vec{k}}$ states. Since they form an orthonormal basis, we have that 

\begin{equation}
\braket{\vec{\mu}',\vec{k}' | \vec{\mu},\vec{k}}=\delta_{\mu'_1,\mu_1}\ldots\delta_{\mu'_n,\mu_n}\delta_{k'_1,k_1}\ldots\delta_{k'_m,k_m}=\delta_{\vec{\mu}',\vec{\mu}}\delta_{\vec{k}',\vec{k}},
\end{equation}
for a graph with $n$ vertices and $m$ edges. The second equality in the previous equation is simply a compact manner to define the product of $n$ and $m$ Kronecker deltas. We shall employ this notation hereafter. Thus, recalling equations (\ref{Hcuk}) and (\ref{HK}), it is readily seen that $$\braket{\vec{\mu}',\vec{k}'|\hat{H}_c|\vec{\mu},\vec{k}}=\frac{1}{4\pi\gamma^3l_{Pl}^2}\sum_iN(v_i)\braket{\vec{\mu}',\vec{k}' | \hat{H}_c^{(i)} | \vec{\mu},\vec{k}},$$ where

\begin{align}
    \braket{\vec{\mu}',\vec{k}' | \hat{H}_c^{(i)} | \vec{\mu},\vec{k}}=&\frac{1}{4\delta_r\delta_\varphi^2}\left(V_{\vec{\mu}+\vec{\delta}_i,\vec{k}}-V_{\vec{\mu}-\vec{\delta}_i,\vec{k}}\right)\left(\delta_{\vec{\mu}',\vec{\mu}+\vec{\delta}_i-\vec{\delta}_{i+1}}+\delta_{\vec{\mu}',\vec{\mu}-\vec{\delta}_i-\vec{\delta}_{i+1}}-\delta_{\vec{\mu}',\vec{\mu}+\vec{\delta}_i+\vec{\delta}_{i+1}}-\delta_{\vec{\mu}',\vec{\mu}-\vec{\delta}_i+\vec{\delta}_{i+1}}\right) \nonumber\\
    &\times\left(\delta_{\vec{k}',\vec{k}-2\vec{k}_i}-\delta_{\vec{k}',\vec{k}+2\vec{k}_i}\right)+\frac{1}{2\delta_r\delta_\varphi^2}\left(V_{\vec{\mu},\vec{k}+2\vec{k}_i}-V_{\vec{\mu},\vec{k}-2\vec{k}_i}\right)\left(\delta_{\vec{\mu}',\vec{\mu}-2\vec{\delta}_i}-2\delta_{\vec{\mu}',\vec{\mu}}+\delta_{\vec{\mu}',\vec{\mu}+2\vec{\delta}_i}\right)\delta_{\vec{k}',\vec{k}} \nonumber\\
    &+\gamma^2H_\Gamma^{(i)}(\vec{\mu},\vec{k})\delta_{\vec{\mu}',\vec{\mu}}\delta_{\vec{k}',\vec{k}}.
		\label{Hp}
\end{align}

Next, the Kronecker deltas are replaced by their integral representation. For this case we have a finite product of deltas, and hence, we may express them as $$\delta_{\vec{\mu}',\vec{\mu}}=\left(\frac{\delta_\varphi}{2\pi}\right)^n\int_0^\frac{2\pi}{\delta_\varphi}d^nb\exp{\left[-i\sum_{i=0}^nb_i(\mu_i-\mu'_i)\right]}, \quad \delta_{\vec{k}',\vec{k}}=\left(\frac{1}{2\pi}\right)^m\int_0^{2\pi}d^mc\exp{\left[-i\sum_{i=0}^nc_i(k_i-k'_i)\right]}.$$ It should be noted that these equations hold only when the values of $\mu_i$ and $k_i$ are being restricted to the super selection sectors. The combination of Kronecker deltas appearing in (\ref{Hp}) yields the following equations,

\begin{align}
\delta_{\vec{\mu}',\vec{\mu}+\vec{\delta}_i-\vec{\delta}_{i+1}}+\delta_{\vec{\mu}',\vec{\mu}-\vec{\delta}_i-\vec{\delta}_{i+1}}& \nonumber \\
-\delta_{\vec{\mu}',\vec{\mu}+\vec{\delta}_i+\vec{\delta}_{i+1}}-\delta_{\vec{\mu}',\vec{\mu}-\vec{\delta}_i+\vec{\delta}_{i+1}}&=-4i\left(\frac{\delta_\varphi}{2\pi}\right)^n\int_0^\frac{2\pi}{\delta_\varphi}d^nb\exp{\left[-i\sum_{i=0}^nb_i(\mu_i-\mu'_i)\right]}\cos(\delta_\varphi b_i)\sin(\delta_\varphi b_{i+1}), \nonumber \\
\delta_{\vec{\mu}',\vec{\mu}-2\vec{\delta}_i}-2\delta_{\vec{\mu}',\vec{\mu}}+\delta_{\vec{\mu}',\vec{\mu}+2\vec{\delta}_i}&=-4\left(\frac{\delta_\varphi}{2\pi}\right)^n\int_0^\frac{2\pi}{\delta_\varphi}d^nb\exp{\left[-i\sum_{i=0}^nb_i(\mu_i-\mu'_i)\right]}\sin^2(\delta_\varphi b_i), \nonumber \\
\delta_{\vec{k}',\vec{k}-2\vec{k}_i}-\delta_{\vec{k}',\vec{k}+2\vec{k}_i}&=-\frac{2i}{(2\pi)^m}\int_0^{2\pi}d^mc\exp{\left[-i\sum_{i=0}^mc_i(k_i-k'_i)\right]}\sin(2c_i).
\label{deltas}
\end{align}
Notice that in the above expressions there is an explicit dependence on the vertix $i$ of the graph. Furthermore, the second line of (\ref{deltas}) depends not only on the vertex $i$, but also on its neighbor labeled as $i+1$. This is a direct consequence of the field aspects of the spherically symmetric model. Another important thing that should be kept in mind is that for each partition there will be integration over $n$ variables in the case of vertices, and $m$ variables in the case of edges. To keep track of this, an additional index will be added to the integration variables, i.e., $b_i^{(j)}$ and $c_i^{(j)}$, and also to $\mu_i^{(j)}$ and $k_i^{(j)}$. We shall adopt the convention in which the upper index $(j)$ refers to the number of partition, where $j=1,2,\ldots,N$, and the lower index $i$ refers to a given vertex or edge. 

After performing the previous steps, the integrand of (\ref{Afi}) can be rearranged as

\begin{align}
    \braket{\vec{\mu}_f,\vec{k}_f | e^{\frac{i}{\hbar}\alpha\hat{H}_c} | \vec{\mu}_i,\vec{k}_i}=\sum_{\vec{\mu}_1,\vec{k}_1,\ldots,\vec{\mu}_{\mathcal{N}-1},\vec{k}_{\mathcal{N}-1}}&\left(\frac{\delta_\varphi}{2\pi}\right)^{n\mathcal{N}}\int_0^\frac{2\pi}{\delta_\varphi}d^nb^{(\mathcal{N})}\ldots d^nb^{(1)} \nonumber \\
    &\times\left(\frac{1}{2\pi}\right)^{m\mathcal{N}}\int_0^{2\pi}d^mc^{(\mathcal{N})}\ldots d^mc^{(1)}e^{\frac{i}{\hbar}S_{eff}}+\mathcal{O}(\varepsilon^2),
		\label{ufi}
\end{align}
with an effective discretized action,

\begin{equation}
    S_{eff}=-\varepsilon\sum_{i,j}\left[\frac{\hbar}{\varepsilon}b_i^{(j)}\left(\mu_i^{(j)}-\mu_i^{(j-1)}\right)+\frac{\hbar}{\varepsilon}c_i^{(j)}\left(k_i^{(j)}-k_i^{(j-1)}\right)+\alpha N(v_i)\mathcal{H}_{eff}^{(i,j)}\right],
		\label{Sdis}
\end{equation}
and 

\begin{align}
\mathcal{H}_{eff}^{(i,j)}=&\frac{1}{2\pi\gamma^3\delta_r\delta_\varphi^2 l_{Pl}^2}\left[\Delta V_1^{(i,j)}\sin\left(\delta_\varphi b_{i+1}^{(j)}\right)\cos\left(\delta_\varphi b_i^{(j)}\right)\sin\left(2c_i^{(j)}\right)+\Delta V_2^{(i,j)}\sin^2\left(\delta_\varphi b_i^{(j)}\right)\right] \nonumber\\
&-\frac{1}{4\pi\gamma l_{Pl}^2}H_\Gamma\left(\mu_i^{(j)},k_i^{(j)}\right),
\end{align}
where

\begin{align}
\Delta V_1^{(i,j)}=&V_{\vec{\mu}+\vec{\delta}_i,\vec{k}}-V_{\vec{\mu}-\vec{\delta}_i,\vec{k}}=4\pi\gamma^{3/2}l_{Pl}^3\left(\left|\mu_i^{(j)}+\delta_\varphi\right|-\left|\mu_i^{(j)}-\delta_\varphi\right|\right)\sqrt{\frac{1}{2}\left|k_i^{(j)}+k_{i-1}^{(j)}\right|}, \nonumber \\
\Delta V_2^{(i,j)}=&V_{\vec{\mu},\vec{k}+2\vec{k}_i}-V_{\vec{\mu},\vec{k}-2\vec{k}_i}=4\pi\gamma^{3/2}l_{Pl}^3|\mu_i|\left(\sqrt{\frac{1}{2}\left|k_i^{(j)}+k_{i-1}^{(j)}+2\right|}-\sqrt{\frac{1}{2}\left|k_i^{(j)}+k_{i-1}^{(j)}-2\right|}\right), \nonumber \\
H_\Gamma\left(\mu_i^{(j)},k_i^{(j)}\right)=&-\frac{1}{2}\left[\frac{1}{\delta_r}\Delta V_2^{(i,j)}\left(1-\Gamma_\varphi^2\left(\mu_i^{(j)},k_i^{(j)}\right)\right)+\frac{1}{\delta_\varphi}\Delta V_1^{(i,j)}\Gamma_\varphi'\left(\mu_i^{(j)},k_i^{(j)}\right)\right]
\end{align}
Note that in equation (\ref{ufi}), the linear approximation has been cast back to an exponential. The action in (\ref{Sdis}) may already be seen as describing an effective theory, albeit at a discrete level. In fact, it is also known as a discrete sum over histories. We are interested, however, in an effective model in the framework of a classical theory, in which the phase space variables are continuous and may take values in the whole real line. In contrast, discrete sums and bounded integrals are featured in equation (\ref{ufi}). The transition between this discrete model to the sought continuous theory is enabled by the use of the generalized Jacobi identity \cite{Kleinert}. It can be expressed as,

\begin{equation}
\sum_{m_1,\ldots,m_n\in\mathbb{Z}}\int_0^{2\pi}f(\theta_1,\ldots,\theta_n,m_1,\ldots,m_n)\exp\left[i\sum_{i=1}^nm_i\theta_i\right]d^n\theta=\int_{\mathbb{R}^{2n}}f(\theta_1,\ldots,\theta_n,x_1,\ldots,x_n)\exp\left[i\sum_{i=1}^nx_i\theta_i\right]d^nxd^n\theta. \nonumber
\end{equation}
In order for this to hold, the function $f$ is required to be continuous, and periodic in the variables $\theta_1,\ldots,\theta_n$. It is clear that this identity is the ideal tool for our purposes, it will change the discrete sums and the bounded integrals for integration over all real values of the involved variables. After applying said identity to the matrix elements of the evolution operator in (\ref{ufi}), we obtain

\begin{align}
    \braket{\vec{\mu}_f,\vec{k}_f | e^{\frac{i}{\hbar}\alpha\hat{H}_c} | \vec{\mu}_i,\vec{k}_i}=&\frac{(\delta_\varphi)^{n\mathcal{N}}}{(2\pi)^{(n+m)\mathcal{N}}}\int_{\mathbb{R}^{2(n+m)\mathcal{N}}}d^n\mu^{(\mathcal{N})}d^mk^{(\mathcal{N})}\ldots d^n\mu^{(1)}d^mk^{(1)}d^nb^{(\mathcal{N})}d^mc^{(\mathcal{N})}\ldots d^nb^{(1)}d^mc^{(1)}e^{\frac{i}{\hbar}S_{eff}} \nonumber \\
		&+\mathcal{O}(\varepsilon^2).
		\label{EvHc}
\end{align}
The $\mu_i^{(j)}$ and $b_i^{(j}$ variables, which were earlier restricted by the super-selection sectors, are now allowed to take any real value. The same applies for $k_i^{(j)}$ and $c_i^{(j)}$, whose values were previously constrained to integers. Despite this, there are still a total of $n\mathcal{N}$ $\mu_i^{(j)}$ and $b_i^{(j}$ variables, one for each vertex and each partition. Similarly for $k_i^{(j)}$ and $c_i^{(j)}$, with $m\mathcal{N}$ variables each, due to the $m$ edges of the graph. The presence of such a number of degrees of freedom is a consequence of the inhomogeneity of the spherically symmetric model.

To deal with the issue related to the high number of variables, the continuum limits $\mathcal{N},n,m\rightarrow\infty$ may now be taken in the path integral calculation. By doing so, one yields what can be thought of as a semiclassical effective model which further resembles a field theory. The previous $n,m\rightarrow\infty$ limits are to be understood as considering a radial graph with infinite bi-valent vertices connected by infinite edges, corresponding thus to a semiclassical state of spherically symmetric quantum geometry. On the other hand, the standard interpretation of the $\mathcal{N}\rightarrow\infty$ is to take infinitesimal (fictitious) time intervals. As a result of applying those limits, we can change the sums over partitions $(j)$ and vertices $i$ to integrals over a time coordinate $\tau$ and a radial coordinate $\tilde{r}$, i.e., $$\sum_{i,j}\rightarrow\int d\tau d\tilde{r}.$$ This procedure also allows us to turn the corresponding vertex or edge index $i$ of the $2(n+m)\mathcal{N}$ integration variables of expression (\ref{EvHc}) into radial dependence, while their partition label $(j)$ turns into temporal dependence. They thus acquire the characteristics of typical field variables, $$\mu_i^{(j)}\rightarrow\mu(\tau,\tilde{r}), \quad b_i^{(j)}\rightarrow b(\tau,\tilde{r}), \quad k_i^{(j)}\rightarrow k(\tau,\tilde{r}), \quad c_i^{(j)}\rightarrow c(\tau,\tilde{r}).$$ Formally, the variables referring to neighboring vertices or edges, for instance $b_{i+1}^{(j)}$ or $k_{i+1}^{(j)}$, should be treated as $b(\tau,\tilde{r}+\Delta\tilde{r})$ and
$k(\tau,\tilde{r}+\Delta\tilde{r})$ for small $\Delta\tilde{r}$. In the continuum limit, the parameter $\Delta\tilde{r}$ becomes arbitrarily small and hence we make $b_{i+1}^{(j)}\rightarrow b(\tau,\tilde{r})$ and $k_{i+1}^{(j)}\rightarrow k(\tau,\tilde{r})$. 

Therefore, after applying the continuum limit we have that

\begin{equation}
    \lim_{\mathcal{N},n,m\rightarrow\infty}\braket{\vec{\mu}_f,\vec{k}_f | e^{\frac{i}{\hbar}\hat{H}_\alpha} | \vec{\mu}_i,\vec{k}_i}=\int\mathcal{D}^n\mu\mathcal{D}^mk\mathcal{D}^n b\mathcal{D}^mce^{\frac{i}{\hbar}S_{eff}},
		\label{DmuDk}
\end{equation}
where the action is given by

\begin{equation}
    S_{eff}=-\int d\tau d\tilde{r}\left(\hbar b\dot{\mu}+\hbar c\dot{k}+\alpha N\mathcal{H}_{eff}\right).
		\label{Seff}
\end{equation}
Here, a dot denotes the derivative with respect to the previously introduced time variable $\tau$. It appears because the limit $\mathcal{N}\rightarrow\infty$ implies $\varepsilon\rightarrow0$, yielding $\tau$ derivatives in the first two terms of (\ref{Sdis}). Additionally in equation (\ref{DmuDk}) we are using the notation due to Feynmann in the context of the path integral framework, this is, $$\int\mathcal{D}\mu=\lim_{N\rightarrow\infty}\prod_{j=1}^N\int d\mu^{(j)}.$$ 

Notice that the variables $\mu$, $k$, $b$, and $c$ are continuous and the phase space spanned by them is $\mathbf{\Gamma}=(c,b;k,\mu)$. Since they are unitless and because $\hbar=l_{Pl}^2$, then, in order for the effective action to have consistent units\footnote{In geometrical units the action has units of squared length.}, $\tilde{r}$ must be unitless as well. The simplectic structure of $\mathbf{\Gamma}$ can be easily read off performing an integration by parts in (\ref{Seff}) and discarding boundary terms, hence obtaining

\begin{equation}
    S_{eff}=\int d\tau dr\left(l_{Pl}\dot{b}\mu+l_{Pl}\dot{c}k-\alpha N\mathcal{H}_{eff}\right),
		\label{Seff2}
\end{equation}
where we defined $r=l_{Pl}\tilde{r}$ as a radial coordinate with units of length and the explicit effective Hamiltonian density reads

\begin{align}
    \mathcal{H}_{eff}=\sqrt{\gamma}&\left[\frac{2}{\gamma^2\delta_\varphi\delta_r}\Delta_1(\mu,k)\sin(\delta_\varphi b)\cos(\delta_\varphi b)\sin(2c)+\frac{2}{\gamma^2\delta_\varphi^2}\Delta_2(\mu,k)\sin^2(\delta_\varphi b)\right. \nonumber \\
		&\left.+\frac{1}{2}\Delta_2(\mu,k)\left(1-\Gamma_\varphi^2(\mu,k)\right)+\frac{1}{2}\Delta_1(\mu,k)\Gamma'_\varphi(\mu,k)\right]
		\label{Heff1}
\end{align}
Furthermore,

\begin{align}
    \Delta_1(\mu,k)=&\frac{\sqrt{|k|}}{\delta_\varphi}\left(|\mu+\delta_\varphi|-|\mu-\delta_\varphi|\right), \quad \Delta_2(\mu,k)=\frac{|\mu|}{\delta_r}\left(\sqrt{\left|k+1\right|}-\sqrt{\left|k-1\right|}\right), \nonumber \\ \Gamma_\varphi(\mu,k)=&-\frac{l_{Pl}k'}{2\delta_\varphi^2}\left(\sqrt{|\mu+\delta_\varphi|}-\sqrt{|\mu-\delta_\varphi|}\right)^2.
\end{align} 
We have thus found an effective Hamiltonian constraint $\mathcal{H}_{eff}$ that includes corrections due to Loop Quantum Gravity effects. Two main types of modifications arise: inverse triad and holonomy corrections. As a result of this, two additional parameters $\delta_r$ and $\delta_\varphi$ are introduced to the model. This Hamiltonian will be analyzed in the next section. 

\section{IV. The Effective Model: Features and Issues}

To enable the comparison between the classical Hamiltonian constraint (\ref{Hc}) and the obtained effective model, the variables $c$, $b$, $k$, and $\mu$ have to be replaced. They can be associated back to the canonical variables $\bar{K}_r$, $\bar{K}_\varphi$, $E^r$ and $E^\varphi$. In fact, one can track down the original variables by examining the flux quantum operators (\ref{E}). The following calculations, done by employing the techniques explained in the last section, are also helpful in the case of the basic holonomy operators, 

\begin{eqnarray}
\Braket{\vec{\mu}',\vec{k}' | \widehat{\sin}\left(\frac{1}{2}\int_{k_i}\bar{K}_rdr\right) | \vec{\mu},\vec{k}}=&-&\left(\frac{\delta_\varphi}{2\pi}\right)^n\int_0^\frac{2\pi}{\delta_\varphi}d^nb\exp{\left[-i\sum_{i=0}^nb_i(\mu_i-\mu'_i)\right]} \nonumber \\
&\times&\frac{1}{(2\pi)^m}\int_0^{2\pi}d^mc\exp{\left[-i\sum_{i=0}^mc_i(k_i-k'_i)\right]}\sin(2c_i) \nonumber \\
\Braket{\vec{\mu}',\vec{k}' | \widehat{\sin}\left(\frac{1}{2}\delta_\varphi\bar{K}_\varphi\right) | \vec{\mu},\vec{k}}=&-&\left(\frac{\delta_\varphi}{2\pi}\right)^n\int_0^\frac{2\pi}{\delta_\varphi}d^nb\exp{\left[-i\sum_{i=0}^nb_i(\mu_i-\mu'_i)\right]}\sin(\delta_\varphi b_i) \nonumber \\
&\times&\frac{1}{(2\pi)^m}\int_0^{2\pi}d^mc\exp{\left[-i\sum_{i=0}^mc_i(k_i-k'_i)\right]}. \nonumber
\end{eqnarray}
Using these matrix elements as a guide, we may then relate the argument of the sine operators in the left-hand side of the past equations to the argument of the sine functions in the right-hand side. In the case of the radial holonomy along the edge $k_i$, the integral can be approximated as $\delta_r\bar{K}_r/2$. This leads us then to rescale the canonical variables as, $$\frac{1}{2}\delta_r\bar{K}_r=2c, \quad \frac{E^r}{\gamma l_{Pl}\delta_r}=k, \quad \frac{1}{2}\bar{K}_\varphi=b, \quad \frac{E^\varphi}{\gamma l_{Pl}}=\mu,$$ where $k$ and $\mu$ are defined such that the simplectic structure of the classical theory (\ref{EstSimp}) coincides with that of the effective model as given by (\ref{Seff2}), up to a common factor.

After implementing these changes the effective action $S_{eff}$ may be expressed in more familiar terms as

\begin{equation}
    S_{eff}=\frac{1}{2}\int d\tau dr\left(\frac{1}{2\gamma}\dot{\bar{K}}_rE^r+\frac{1}{\gamma}\dot{\bar{K}}_\varphi E^\varphi-N\mathcal{H}_{eff}\right).
		\label{Seff3}
\end{equation}
The multiplicative constant outside the integral sign is unimportant and also, additional factors were absorbed into the lapse function $N$. Now we can write the effective Hamiltonian constraint in a way that resembles the classical one, namely, 

\begin{align}
    \mathcal{H}_{eff}=&-\frac{1}{\gamma^2\sqrt{|E^r|}}\left[\beta_2(\bar{K}_\varphi)\beta_3(\bar{K}_r)\alpha_2(E^\varphi)E^r+\frac{1}{2}\beta_1^2(\bar{K}_\varphi)\alpha_1(E^r)E^\varphi\right] \nonumber \\
    &-\frac{E^\varphi}{2\sqrt{|E^r|}}(1-\Gamma_{\varphi eff}^2)\alpha_1(E^r)-\sqrt{|E^r|}\alpha_2(E^\varphi)\Gamma'_{\varphi eff}.
\end{align}
With correction functions given by,

\begin{align}
    \beta_1(\bar{K}_\varphi)=&\frac{2}{\delta_\varphi}\sin\left(\frac{1}{2}\delta_\varphi\bar{K}_\varphi\right), \quad \beta_2(\bar{K}_\varphi)=\frac{2}{\delta_\varphi}\sin\left(\frac{1}{2}\delta_\varphi\bar{K}_\varphi\right)\cos\left(\frac{1}{2}\delta_\varphi\bar{K}_\varphi\right), \quad \beta_3(\bar{K}_r)=\frac{2}{\delta_r}\sin\left(\frac{1}{2}\delta_r\bar{K}_r\right), \nonumber \\
    \alpha_1(E^r)=&\frac{\sqrt{|E^r|}}{\bar{\delta}_r^2}\left(\sqrt{\left|E^r+\bar{\delta}_r^2\right|}-\sqrt{\left|E^r-\bar{\delta}_r^2\right|}\right), \quad \alpha_2(E^\varphi)=\frac{1}{2\bar{\delta}_\varphi}\left(\left|E^\varphi+\bar{\delta}_\varphi\right|-|E^\varphi-\bar{\delta}_\varphi|\right), \nonumber \\ 
		\Gamma_{\varphi eff}=&-\frac{E^{r\prime}}{2\bar{\delta}_\varphi^2}\left(\sqrt{\left|E^\varphi+\bar{\delta}_\varphi\right|}-\sqrt{\left|E^\varphi-\bar{\delta}_\varphi\right|}\right)^2.
		\label{hg}
\end{align}
Here, $\bar{\delta}_\varphi=\gamma l_{Pl}\delta_\varphi$ and $\bar{\delta}_r^2=\gamma\delta_r^2$. Moreover, since the radial holonomy parameter is expected to be of the order of the Planck length, we have set $\delta_r=l_{Pl}$ (the notation $\delta_r$ will be preferred hereafter as it emphasizes the role of the quantity within the correction functions). The $\beta$ functions are modifications due to holonomies and depend on extrinsic curvature $\bar{K}_r$ and $\bar{K}_\varphi$ , while the $\alpha$ and $\Gamma_{\varphi eff}$ functions are inverse triad corrections that, as the name suggests, involve the densitized triad components $E^r$ and $E^\varphi$. It can be verified that when applying the classical limits, $\delta_\varphi\bar{K}_\varphi,\delta_r\bar{K}_r,\bar{\delta}_\varphi/E^\varphi,\bar{\delta}_r^2/E^r\ll1$, then

\begin{equation}
    \beta_{1,2}(\bar{K}_\varphi)\rightarrow\bar{K}_\varphi, \quad \beta_3(\bar{K}_r)\rightarrow\bar{K}_r, \quad \alpha_{1,2}\rightarrow1, \quad \Gamma_{\varphi eff}\rightarrow\Gamma_{\varphi}.  
\end{equation}
The previous limits are expected to hold in spacetime regions of low curvature, where quantum effects can be neglected and the semiclassical model reduces to its classical version. Indeed, it is readily seen that in this so-called classical limit, we have that $\mathcal{H}_{eff}\rightarrow\mathcal{H}_c$ as given by equation (\ref{Hc}). 

For effective spherical symmetry models, another key property that should be analyzed is their general covariance, i.e., the invariance under diffeomorphisms of its modified action. In the classical theory this property is encoded in the algebra of constraints \cite{Thiemann_2007}, e.g, in (\ref{CAEsf}) for this case. It is a necessary, but not sufficient, condition for diffeomorphism invariance. Hence, the model will be generally covariant only if the obtained effective Hamiltonian constraint satisfies a similar algebra. Unfortunately one can verify that when all the derived corrections are present, the structure of the classical algebra breaks down,

\begin{equation}
\{H_{eff}[N],C_c[N_r]\}\neq H_{eff}[N_rN'], \quad \{H_{eff}[N],H_{eff}[M]\}\neq C_c\left[\frac{E^r}{(E^\varphi)^2}\left(MN'-NM'\right)\right].
\label{CAEff}
\end{equation}
Not only that, but an explicit (and cumbersome) calculation of these Poisson brackets reveals that there appear anomalous terms, these are, terms that prevent the algebra from being closed. Covariance, thus, is violated in this complete effective model. 

\subsection{IV.A. Simplified Model}

The corrections responsible for the appearance of anomalous terms in the algebra of constraints are those related to the $\bar{K}_r$ and $E^\varphi$ variables, i.e., the $\alpha_2(E^\varphi)$, the $\beta_3(\bar{K}_r)$, and the $\Gamma_{\varphi eff}$ modification functions in (\ref{hg}). For this reason, a simplified effective model can be considered by neglecting the problematic corrections and keeping only those that do not alter the closure property of the algebra. This would correspond to the case of choosing arbitrarily small radial intervals $I$ for the integration region of the holonomy $h_r$ in equation (\ref{hr}) and the subsequent fluxes $E(S_{r\theta})$, $E(S_{r\varphi})$. The character of these quantities is non-local, as opposed to the point-holonomies $h_{\theta,\varphi}$ and the radial flux $E(S_{\theta\varphi})$ \cite{brahma}. The quantum corrections due to these variables can be made negligible and therefore discarded from the effective analysis. Consequently the algebra remains first class and the corrected constraints still generate gauge transformations in phase space. The simplified modifications of the Hamiltonian constraint can thus be expressed as

\begin{align}
H_{eff}[N]=&-\int drN\left[\frac{1}{\gamma^2\sqrt{|E^r|}}\left(\beta_2(\bar{K}_\varphi)\bar{K}_rE^r+\frac{1}{2}\beta_1^2(\bar{K}_\varphi)\alpha_1(E^r)E^\varphi\right)\right. \nonumber \\
&\hspace{2cm}\left.+\frac{E^\varphi}{2\sqrt{|E^r|}}(1-\Gamma_\varphi^2)\alpha_1(E^r)+\sqrt{|E^r|}\Gamma'_\varphi\right].
\label{HeffInv}
\end{align}
This will be the final form of our effective spherical model. The classical limit corresponds to $\delta_\varphi,\bar{\delta}_r\rightarrow0$, which makes $\alpha_1(E^r)\rightarrow1$ and $\beta_{1,2}(\bar{K}_\varphi)\rightarrow\bar{K}_\varphi$. It can then be verified that the algebra of constraints of the simplified model is anomaly-free and preserves the classical structure,
		
\begin{align}
    &\{C_c[N_r],H_{eff}[N]\}=H_{eff}[N_rN'], \quad \{C_c[N_r],C_c[M_r]\}=C_c[N_rM'_r-M_rN'_r], \nonumber \\
    &\{H_{eff}[N],H_{eff}[M]\}=C_c\left[\frac{d\beta_2}{d\bar{K}_\varphi}\frac{E^r}{(E^\varphi)^2}\left(NM'-MN'\right)\right].
		\label{CAEffSimp}
\end{align}
Note however that the last bracket acquires a modification factor due to the presence of loop quantum corrections. This is sometimes referred to as a deformed algebra (compare with the classical algebra of (\ref{CAEsf})). 

At this point the election of the modified angular closed holonomy $\hat{\bar{h}}_{\theta\varphi}$ leading to equation (\ref{HKmod}) can be explained. If one completes the quantization procedure utilizing the angular holonomy with the usual parameter, namely, $$\hat{h}_{\theta\varphi}=\hat{h}_\theta\left(v_i,\delta_\varphi\right)\hat{h}_\varphi\left(v_i,\delta_\varphi\right)\hat{h}_\theta^{-1}\left(v_i,\delta_\varphi\right)\hat{h}_\varphi^{-1}\left(v_i,\delta_\varphi\right),$$ instead of that given by (\ref{htpmod}), then a correction function $\beta_1^{(0)}=\sin(\delta_\varphi\bar{K}_\varphi)/\delta_\varphi$ at the effective level is obtained as a result of applying the path integral method. The function $\beta_2$ would remain unchanged in this case. However, a Hamiltonian constraint with both $\beta_1^{(0)}$ and $\beta_2$ as holonomy modifications does not yield a first class algebra \cite{Tibrewala_2012}, this automatically implies the loss of covariance, even when a simplified version of the model is considered. It is interesting to realize that the holonomies along closed paths chosen to perform the loop quantization scheme can have contrasting consequences for their respective semiclassical counterparts. 

We can now proceed to analyze the effective dynamics of this model while being certain that gauge transformations are still part of the semiclassical theory. 

\section{V. The Effective Field Equations}

In this section we explicitly solve the effective dynamics of the simplified model in different gauges and give a physical interpretation to the corresponding solutions. The semiclassical field equations we need for this purpose are given by the constraints, along with the following Hamilton equations, $$\mathcal{C}_c=0, \quad \mathcal{H}_{eff}=0, \quad \frac{\delta H_T^{eff}}{\delta\bar{K}_r}=-\frac{1}{2\gamma}\dot{E}^r, \quad \frac{\delta H_T^{eff}}{\delta\bar{K}_\varphi}=-\frac{1}{\gamma}\dot{E}^\varphi, \quad \frac{\delta H_T^{eff}}{\delta E^\varphi}=\frac{1}{\gamma}\dot{\bar{K}}_\varphi.$$ Here $H_T^{eff}=H_{eff}[N]+C_c[N_r]$, with $H_{eff}[N]$ given by the simplified effective Hamiltonian constraint (\ref{HeffInv}). These equations turn respectively into

\begin{align}
&2E^\varphi\bar{K}'_\varphi-(E^r)'\bar{K}_r=0, \quad \frac{1}{\gamma^2}\left(\beta_2\bar{K}_rE^r+\frac{1}{2}\beta_1^2\alpha_1E^\varphi\right)+\frac{E^\varphi}{2}(1-\Gamma_\varphi^2)\alpha_1+E^r\Gamma'_\varphi=0, \nonumber \\
&\dot{E}^r=\frac{2}{\gamma}N\beta_2\sqrt{\left|E^r\right|}+N_r(E^r)', \quad \dot{E}^\varphi=\frac{N}{\gamma}\left[\frac{E^\varphi\alpha_1}{2\sqrt{\left|E^r\right|}}\frac{d\beta_1^2}{d\bar{K}_\varphi}+\bar{K}_r\sqrt{\left|E^r\right|}\frac{d\beta_2}{d\bar{K}_\varphi}\right]+\left(N_rE^\varphi\right)', \nonumber \\ 
&\dot{\bar{K}}_\varphi=\frac{1}{2\gamma(E^\varphi)^2\sqrt{E^r}}\left[\frac{\gamma^2}{2}(E^r)'\left[(NE^r)'+N'E^r\right]-N\alpha_1\left([\gamma^2+\beta_1^2][E^\varphi]^2+\frac{1}{4}\left[\gamma(E^r)'\right]^2\right)\right]+N_r\bar{K}'_\varphi.
\label{EFE}
\end{align}
In the past equations the variable dependence of the functions $\alpha_1(E^r)$ and $\beta_{1,2}(\bar{K}_\varphi)$ has been dropped in order to lighten notation. Also, while there are already available explicit forms of these quantum modifications, it is more convenient to treat them as unspecified functions for the time being. In order to solve the effective equations particular gauges will be chosen guided by the intuition given by classical solutions of GR. Based on this, one expects to find solutions that represent quantum corrected versions of the Schwarzschild black hole whose exterior and interior region can both be described by the spherically symmetric model. Since there are two constraints, there is gauge freedom to initially fix two canonical variables or Lagrange multipliers during the calculation process. The election of the gauge will determine which region of the black hole is being analyzed.

\subsection{V.A. The Exterior Region}

The exterior region of the black hole is characterized by an inhomogeneous gauge. In such a gauge, there is no time dependence on any of the fields, e.g., $\dot{N}=\dot{E}^r=\dot{E}^\varphi=0$. Additionally the areal radius can be chosen so that $E^r=r^2$, and the static condition $N_r=0$ can be imposed. This will be called the Schwarzschild gauge. The $\dot{E}^r$ effective field equation in (\ref{EFE}) implies then that $$\beta_2=\frac{2}{\delta_\varphi}\sin\left(\frac{1}{2}\delta_\varphi\bar{K}_\varphi\right)\cos\left(\frac{1}{2}\delta_\varphi\bar{K}_\varphi\right)=\frac{1}{\delta_\varphi}\sin\left(\delta_\varphi\bar{K}_\varphi\right)=0$$ in this gauge. Hence, $\bar{K}_\varphi=n\pi/\delta_\varphi$ with $n\in\mathbb{N}$ and $\beta_1=2\sin\left(n\pi/2\right)/\delta_\varphi$. Using the diffeomorphism constraint it is immediate to see that $\bar{K}_r=0$, which in turn allows for the effective Hamiltonian constraint to be rearranged as

\begin{equation}
(1+\gamma_n^2)\left(\frac{E^\varphi}{r}\right)^3-\frac{E^\varphi}{r}+\frac{2r}{\alpha_1}\left(\frac{E^\varphi}{r}\right)'=0,
\label{En}
\end{equation}
where $$\gamma_n^2=\frac{4}{\gamma^2\delta_\varphi^2}\sin^2\left(\frac{n\pi}{2}\right)=\begin{cases}
    0, &\text{for even $n$,} \\
    4/\gamma^2\delta_\varphi^2, &\text{for odd $n$.} 
    \end{cases}$$ The differential equation (\ref{En}) can be easily solved, yielding 

\begin{equation}
\left(\frac{E^\varphi}{r}\right)^2=E_n^2=\frac{1}{1+\gamma_n^2-2\bar{m}e^{-R(\bar{r})}}, \quad R(\bar{r})=\int\frac{\alpha_1}{\bar{r}}d\bar{r}.
\label{Rr}
\end{equation}
Here $\bar{r}=r/\bar{\delta}_r$ is a scaled radial coordinate and $\bar{m}=m/\bar{\delta}_r$ is an integration constant chosen so that it coincides with the mass $m$ of the black hole in the classical limit. The missing Lagrange multiplier $N$ can be found by examining the $\dot{\bar{K}}_\varphi$ field equation in (\ref{EFE}). Inserting all of the previous results we obtain $$2rN'+\left[2-\alpha_1\left((1+\gamma_n^2)E_n^2+1\right)\right]N=0.$$ The solution for $N$ reads

\begin{equation}
N^2=\frac{e^{2R(\bar{r})}}{\bar{r}^2}\left(1+\gamma_n^2-2\bar{m}e^{-R(\bar{r})}\right).
\end{equation}
One can identify the result of including loop quantum effects in the above solutions. Holonomy modifications $\beta_{1,2}(\bar{K}_\varphi)$ are responsible for the appearance of the $\gamma_n$ factor, while the inverse triad correction $\alpha_1(E^r)$ yields the non-trivial function $e^{R(\bar{r})}$. The effective solution without inverse triad correction, i.e., with $\alpha_1=1$, was previously found and studied in \cite{brahma}.

It is straightforward to realize that the classical limit corresponds to the case $\gamma_0=0$ (or equivalently even $n$) and $\bar{\delta}_r\rightarrow0$. Since $\alpha_1$ is a correction function such that $\lim_{\bar{\delta}_r\rightarrow0}\alpha_1=1$, then $e^{R(\bar{r})}\rightarrow\bar{r}$ as $\bar{\delta}_r\rightarrow0$, and the classical Schwarzschild solution is recovered, i.e., $$E_0^2\rightarrow\frac{1}{1-2m/r}, \quad \bar{N}^2\rightarrow1-\frac{2m}{r}.$$ Note that by looking at the explicit solutions for $N^2$ and $E_n^2$ one can realize that they are valid so long as $e^{R(\bar{r})}>2\bar{m}/(1+\gamma_n^2)$. In the classical limit this is $r>2m$, which corresponds to the exterior region of the black hole. At this point it is relevant to discuss the interpretation of the parameter $\gamma_n$. It was already mentioned that in order to return to the classical Schwarzschild solution, even values of $n$ have to be considered. On the other hand, if $\delta_\varphi\ll1$ for the holonomy parameter, odd values of $n$ lead to a large $\gamma_n$. This can be seen as describing strong quantum effects. Such effects however are not expected to arise in the exterior region of a black hole. Thus, a consistent physical picture of a black hole with loop quantum modifications may be described by $N$ and $E_0^2$, where the function $e^{R(\bar{r})}$ has a small parameter $\bar{\delta}_r$. On this matter, the explicit $\alpha_1$ correction in this gauge can be expressed as 

\begin{equation}
\alpha_1=\bar{r}\left(\sqrt{\left|\bar{r}^2+1\right|}-\sqrt{\left|\bar{r}^2-1\right|}\right).
\label{alpha1}
\end{equation} 
If this correction is inserted in (\ref{Rr}), the exponential becomes

\begin{equation}
    e^{R(\bar{r})}=\begin{cases} \displaystyle
    \frac{1}{2}e^{(\alpha_1-1)/2}\sqrt{\left(\bar{r}+\sqrt{\bar{r}^2+1}\right)\left(\bar{r}+\sqrt{\bar{r}^2-1}\right)}, & \bar{r}\geq1, \\
    \displaystyle\frac{1}{2}\sqrt{\bar{r}+\sqrt{\bar{r}^2+1}}\exp\left[\frac{1}{2}\left(\alpha_1-1-\arctan\left(\frac{\bar{r}}{\sqrt{1-\bar{r}^2}}\right)+\frac{\pi}{2}\right)\right], & 0\leq\bar{r}<1.
\end{cases}
\label{eRr}
\end{equation}
Integration constants were chosen so that $e^{R(\bar{r})}$ is continuous. The last part of this piece-wise function is included just for completeness. In the exterior of a physical black hole the following inequality should hold $e^{R(\bar{r})}>2\bar{m}>1$. The values $0\leq\bar{r}<1$ are then not expected to fall within the domain of this solution. Inverse triad corrections with their corresponding effective solution, but without holonomy modifications, were studied in the same context in \cite{Tibrewala_2012}. The analysis here can be seen as a generalization which includes both holonomy and inverse triad effects.

\subsection{V.B. The Interior Region}

We turn now to the interior of the black hole. Inside it, the time and radial coordinates change their causal character and thus, the region can be described as a cosmological model. Contrary to its exterior counterpart, this portion of space-time is characterized by a homogeneous gauge in which none of the fields depend on the radial variable, for example, $N'=(E^r)'=(E^\varphi)'=0$. In analogy to the inhomogeneous case one can choose the areal radius as $E^r=t^2$ and impose the static condition $N_r=0$. We will refer to this as the Kantowski-Sachs gauge. The steps followed to find the interior solution are similar to those of the past subsection.

By examining the $\dot{E}^r$ effective field equation in (\ref{EFE}) one finds that in this gauge, 

\begin{equation}
\beta_2=\frac{\gamma}{N}.
\label{g2}
\end{equation}
Deriving this with respect to $t$ allows us to write an expression for $\dot{\beta}_2=\dot{\bar{K}}_\varphi d\beta_2/d\bar{K}_\varphi$ and further use it in the Hamilton equation for $\dot{\bar{K}}_\varphi$. Doing so yields

\begin{equation}
\dot{N}=\frac{N^3\alpha_1}{2t}\left(1+\frac{\beta_1^2}{\gamma^2}\right)\frac{d\beta_2}{d\bar{K}_\varphi},
\end{equation}
Additionally, because $\beta_{1,2}$ are trigonometric functions, it can be seen by combining the functional form of $\beta_2$ and (\ref{g2}) that $$\beta_1^2=\frac{2}{\delta_\varphi^2}\left(1-\sqrt{1-\frac{\gamma^2\delta_\varphi^2}{N^2}}\right), \quad \frac{d\beta_2}{d\bar{K}_\varphi}=\sqrt{1-\frac{\gamma^2\delta_\varphi^2}{N^2}}.$$ A convenient change of variable $u^2=1-\gamma^2\delta_\varphi^2/N^2$ simplifies the process of solving the previous differential equation for $\dot{N}$. Here we only show the final result,

\begin{equation}
    N^2=\frac{1}{\left[2\bar{m}e^{-T(\bar{t})}-1\right]\left[1-\gamma^2\delta_\varphi^2\left(2\bar{m}e^{-T(\bar{t})}-1\right)/4\right]}, \quad T(\bar{t})=\int\frac{\alpha_1}{\bar{t}}d\bar{t}.
		\label{Nint}
\end{equation}
As in the previous exterior solution, we have introduced a scaled time coordinate $\bar{t}=t/\bar{\delta}_r$ and $\bar{m}$ is the scaled mass of the black hole. To obtain a solution for $E^\varphi$ we need to consider first the $\dot{E}^\varphi$ field equation, which can be written as $$\bar{K}_r\frac{d\beta_2}{d\bar{K}_\varphi}=\frac{\gamma\dot{E}^\varphi}{tN}-\frac{E^\varphi\alpha_1}{2t^2}\frac{d\beta_1^2}{d\bar{K}_\varphi},$$ with $d\beta_1^2/d\bar{K}_\varphi=2\gamma/N$ due to trigonometric identities. Inserting $\bar{K}_r$ into the effective Hamiltonian constraint of the equations in (\ref{EFE}) then gives $$t\frac{d}{dt}\left(\frac{E^\varphi}{t}\right)+\left[1-\alpha_1+\frac{N^2\alpha_1}{2}\left(1+\frac{\beta_1^2}{\gamma^2}\right)\frac{d\beta_2}{d\bar{K}_\varphi}\right]\frac{E^\varphi}{t}=0,$$ whose solution is

\begin{equation}
    \left(\frac{E^\varphi}{t}\right)^2=E^2=\frac{e^{2T(\bar{t})}}{\bar{t}^2}\left[2\bar{m}e^{-T(\bar{t})}-1\right]\left[1-\frac{\gamma^2\delta_\varphi^2}{4}\left(2\bar{m}e^{-T(\bar{t})}-1\right)\right].
		\label{Eint}
\end{equation}

In a similar way as in the exterior region, loop quantum modifications can be linked to particular parts of the past expressions. Holonomy effects are related to the factor $\gamma^2\delta_\varphi^2\left(2\bar{m}e^{-T(\bar{t})}-1\right)/4$ appearing in $N^2$ and $E^2$, whereas inverse triad corrections are contained exclusively in $e^{T(\bar{t})}$. The explicit form of this exponential is completely analogous to (\ref{Rr}), namely,
	
\begin{equation}
    e^{T(\bar{t})}=\begin{cases} \displaystyle
    \frac{1}{2}e^{(\alpha_1-1)/2}\sqrt{\left(\bar{t}+\sqrt{\bar{t}^2+1}\right)\left(\bar{t}+\sqrt{\bar{t}^2-1}\right)}, & \bar{t}\geq1, \\
    \displaystyle\frac{1}{2}\sqrt{\bar{t}+\sqrt{\bar{t}^2+1}}\exp\left[\frac{1}{2}\left(\alpha_1-1-\arctan\left(\frac{\bar{t}}{\sqrt{1-\bar{t}^2}}\right)+\frac{\pi}{2}\right)\right], & 0\leq\bar{t}<1.
\end{cases}
\end{equation}
When applying the classical limit $\delta_\varphi,\bar{\delta}_r\rightarrow0$ to the above solutions, the interior Schwarzschild space-time is recovered, this is, $$N^2\rightarrow\frac{1}{2m/t-1}, \quad E^2\rightarrow\frac{2m}{t}-1.$$ Finally, since $N^2,E^2\geq0$ has to hold in order for the interior solution to be well-defined, an analysis of the roots of (\ref{Eint}) shows that its domain is 

\begin{equation}
\frac{2\gamma^2\delta_\varphi^2\bar{m}}{(4+\gamma^2\delta_\varphi^2)}<e^{T(\bar{t})}<2\bar{m}. 
\label{intdom}
\end{equation}
Of course, in the classical limit one recovers the well-known interior domain of the Schwarzschild black hole, $0<t<2m$. Note that due to holonomy corrections, the singularity that would classically appear at $t=0$ is excluded from the effective interior.

\subsection{V.C. The Painlev\'e-Gullstrand Gauge}

Another interesting gauge to consider is that defined by the use of the Painlev\'e-Gullstrand (PG) coordinates. Because they are able to penetrate the horizon of the Schwarzschild space-time, such coordinates turn out to be extremely helpful for the description of the black hole. Naturally, PG coordinates cover both the exterior and the interior region. In this coordinate system the usual Schwarzschild time variable $t$ is changed to $t_{PG}=t+F(r)$ so that it can be interpreted as depicting space-time as seen from an observer in free fall into the black hole. In fact, the function $F(r)$ is defined so that in slices of constant $t_{PG}$, the induced spatial metric is flat. This condition imposes then that $(E^\varphi)^2=E^r=r^2$ in the PG gauge (see the space-time metric (\ref{ds2Esf})). To avoid confusion between the time coordinate $t$ previously employed in this section, we are denoting the Painlev\'e-Gullstrand time as $t_{PG}$.

In the following, we will search for an effective solution in the PG gauge, i.e., using $(E^\varphi)^2=E^r=r^2$ and the time-independence condition $\dot{N}=\dot{E}^r=\dot{E}^\varphi=0$ applied to all variables. Here the dot denotes derivation with respect to $t_{PG}$. The first step is to find $\beta_2$ on-shell from the effective Hamilton equation for $\dot{E}^r$. This yields $$\beta_2=-\gamma\frac{N_r}{N}.$$ It immediate to see from the diffeomorphism constraint that $\bar{K}_r=\bar{K}'_{\varphi}$. Using these two past results and after some algebraic steps, the $\dot{E}^\varphi$ field equation can be expressed as 

\begin{equation}
rN'=\left(\alpha_1-1\right)N.
\label{NPG}
\end{equation}
It is straightforward to find that $N=e^{R(\bar{r})}/r$ is a solution to (\ref{NPG}), where $e^{R(\bar{r})}$ was previously defined in (\ref{Rr}) and an explicit form of it is given by equation (\ref{eRr}). Next, consider the effective dynamics equation for $\dot{\bar{K}}_\varphi$, which in this gauge becomes 

\begin{equation}
r\bar{K}'_\varphi\frac{N_r}{N}=\frac{\alpha_1\beta_1^2}{2\gamma},
\label{NrEq}
\end{equation} 
where using trigonometric identities we have that $$\bar{K}'_\varphi=\frac{-\gamma}{\sqrt{1-\gamma^2\delta_\varphi^2N_r^2/N^2}}\left(\frac{N_r}{N}\right)', \quad \beta_1^2=\frac{2}{\delta_\varphi^2}\left(1-\sqrt{1-\gamma^2\delta_\varphi^2\frac{N_r^2}{N^2}}\right).$$ Note that (\ref{NrEq}) is a differential equation for $N_r/N$. The easiest way to solve it is by applying the following change of variable $u^2=1-\gamma^2\delta_\varphi^2N_r^2/N^2$, which then leads to

\begin{equation}
N_r=\frac{1}{\bar{r}}\sqrt{2\bar{m}e^{R(\bar{r})}\left(1-\frac{1}{2}\bar{m}\gamma^2\delta_\varphi^2e^{-R(\bar{r})}\right)}.
\label{Nr}
\end{equation}
Here the scaled mass of the black hole $\bar{m}$ appears as an integration constant. The solution with holonomy effects only was previously described in \cite{asier}.

One can readily verify that in the classical limit, $\delta_\varphi,\bar{\delta}_r\rightarrow0$, the lapse and radial shift become $N\rightarrow1$ and $N_r\rightarrow\sqrt{2m/r}$, respectively. This describes the Schwarzschild metric in PG coordinates $$ds^2=-dt_{PG}^2+\left(dr+\sqrt{\frac{2m}{r}}dt_{PG}\right)^2+r^2d\Omega^2.$$ The domain of this solution is restricted by values such that $N_r\in\mathbb{R}$ in equation (\ref{Nr}), and hence $e^{R(\bar{r})}\geq\gamma^2\delta_\varphi^2\bar{m}/2$. Classically this corresponds to the well-known property that PG coordinates cover the exterior and interior region of the black hole, i.e., $r\geq0$.

\subsection{V.D. A Dirac Observable}

So far, solutions for the simplified effective model in three different gauges have been presented. In this subsection we will introduce a Dirac observable (a gauge invariant quantity) for the Hamiltonian system described by the effective constraint in (\ref{HeffInv}). Since first class constraints generate gauge transformations through Poisson brackets, a Dirac observable $\mathscr{O}$ can be defined as a phase space function whose Poisson bracket with each of the constraints identically vanishes. It is also possible that said brackets vanish only in the constraint hypersurface $S$, in which case, $\mathscr{O}$ is called a weak Dirac observable. In what follows and for simplicity, we will not distinguish between any of the previous types of situations. 

In the canonical analysis of the Schwarzschild space-time, it is well-known that the mass $m$ of the black hole is a Dirac observable. In terms of the phase space variables used in this paper, this quantity is given by

\begin{equation}     
\mathscr{O}=\frac{1}{2}\sqrt{|E^r|}\left[1+\frac{\bar{K}_\varphi^2}{\gamma^2}-\left(\frac{(E^r)'}{2E^\varphi}\right)^2\right].
\label{OD}
\end{equation}
It can be easily verified that in any gauge, or equivalently any coordinate system used to describe the metric, $\mathscr{O}=m$. Hence, the mass of the black hole is a physical observable. Based on $\mathscr{O}$, the analogous phase space function for our simplified effective model can be heuristically found. Note that the loop quantum corrected Hamiltonian constraint (\ref{HeffInv}) can be obtained from the classical one (\ref{Hc}) by changing $$\frac{1}{\sqrt{\left|E^r\right|}}\rightarrow\frac{\alpha_1(E^r)}{\sqrt{\left|E^r\right|}}, \quad \bar{K}_\varphi\rightarrow\beta_2(\bar{K}_\varphi), \quad \bar{K}^2_\varphi\rightarrow\beta_1^2(\bar{K}_\varphi).$$ Additionally, we can use the previous solutions as a guide in order to express the mass as a phase space function. One is then led to propose the quantity,

\begin{equation}
    \mathscr{O}_{eff}=\frac{1}{2}e^{R(E^r)}\left[1+\frac{\beta_1^2(\bar{K}_\varphi)}{\gamma^2}-\left(\frac{(E^r)'}{2E^\varphi}\right)^2\right], \quad \frac{dR(E^r)}{dE^r}=\frac{\alpha_1}{2E^r},
\end{equation}
as a possible Dirac observable for the effective model. In fact, a direct computation of the Poisson brackets with the respective constraints confirms this intuition, i.e., 

\begin{eqnarray}
    \left\{\mathscr{O}_{eff},H_{eff}[N]\right\}&=&-C_c\left[\frac{N\sqrt{|E^r|}\left(E^r\right)'}{2\left(E^\varphi\right)^3}\frac{d\beta_2}{d\bar{K}_\varphi}e^{R(E^r)}\right], \nonumber \\ \left\{\mathscr{O}_{eff},C_c[N_r]\right\}&=&C_c\left[\frac{N_r\beta_2}{4\gamma^2E^\varphi}e^{R(E^r)}\right]-H_{eff}\left[\frac{N_r\left(E^r\right)'}{2E^\varphi\sqrt{|E^r|}}e^{R(E^r)}\right].
\end{eqnarray}
Thus, $$\left.\left\{\mathscr{O}_{eff},H_{eff}[N]\right\}\right|_S=0, \quad \left.\left\{\mathscr{O}_{eff},C_c[N_r]\right\}\right|_S=0,$$ and $\mathscr{O}_{eff}$ is indeed a Dirac observable. Inserting the previously found solutions of this section yields that $\mathscr{O}_{eff}=\bar{m}$. This further shows that we are dealing with a Hamiltonian system in which consistent gauge transformations can be applied and a physical observable exists.

\section{VI. Geometric Interpretation of the Effective Solutions}

In the classical case, solutions obtained from the Hamiltonian formalism of General Relativity can be interpreted as coefficients of a space-time metric. In turn, the metric can be given certain physical meaning and relevant predictions can be made based on the particular geometry it describes. It would be tempting then to treat solutions from the modified theory in this same manner, yielding an effective metric that includes Loop Quantum Gravity corrections within itself. However, in order to do so, the issue of covariance becomes relevant as already mentioned in section IV. The requisite of a closed algebra of constraints was introduced there, so that gauge transformations were still presented in the new theory and in fact, generated by the modified constraints. While this is a necessary condition, it is not sufficient. It needs to be further verified that said gauge transformations correspond to those induced by an infinitesimal change of coordinates \cite{zhang, brahma, belfaqih}, i.e., to the Lie derivative of the metric along the infinitesimal vector that generates such change. The covariance analysis of holonomy corrections in the effective model found here has already been carried out in appendix A of \cite{asier}. In the following we provide complementary calculations that further support the arguments given in said reference. 

It turns out that in general, the fields found as solutions to the effective field equations do not satisfy the sufficient condition for covariance on their own. Recent schemes suggest that in addition to the effective solutions, extra correction factors that depend on the modification functions of the Hamiltonian need to be included to obtain quantities that transform covariantly and therefore, can be interpreted as metric coefficients. This construction is sometimes called an emergent metric \cite{belfaqih}, denoted hereafter as $\bar{g}_{\mu\nu}$. At this point, the structure functions of the algebra of constraints become relevant. In the classical spherically symmetric model, the function that appears in the $\{H_c[N],H_c[M]\}$ bracket of (\ref{CAEsf}) corresponds to a component of the inverse spatial metric $q^{ab}$, namely, $q^{rr}=E^r/(E^\varphi)^2$. This comes as no surprise since the algebra of constraints coincides with the algebra of hyper-surface deformations, which depends on the foliation employed in the ADM decomposition of space-time. Examining the algebra of effective constraints (\ref{CAEffSimp}) it is natural then to consider an emergent metric with inverse component, 

\begin{equation}
\bar{q}^{rr}=\frac{d\gamma_2}{dK_\varphi}\frac{E^r}{(E^\varphi)^2},
\label{bqrr}
\end{equation}
and leave the rest of the metric coefficients unmodified. It has also been shown \cite{Tibrewala_2012} that another possibility is to incorporate the extra correction into a new lapse $\bar{N}$.

Let us verify thereby if the $\bar{q}_{rr}$ component implied by (\ref{bqrr}) and the corrected lapse $\bar{N}$ alternative transform covariantly. Consider the following general spherically symmetric line element, $$ds^2=-\bar{N}^2dt^2+\bar{q}_{rr}\left(dr+N_rdt\right)^2+E^rd\Omega^2.$$ Under an infinitesimal change of coordinates $x^\mu\rightarrow x^\mu+\xi^\mu$, the metric quantities of interest transform as $$\bar{q}_{rr}\rightarrow\bar{q}_{rr}+\delta_\xi\bar{q}_{rr}, \quad \bar{N}\rightarrow\bar{N}+\delta_\xi\bar{N},$$ with

\begin{equation}
\delta_\xi\bar{q}_{rr}=\dot{\bar{q}}_{rr}\xi^0+\bar{q}'_{rr}\xi^1+2\bar{q}_{rr}\left[N_r\left(\xi^0\right)'+\left(\xi^1\right)'\right], \quad \delta_\xi\bar{N}=\left(\bar{N}\xi^0\right)\dot{}+\bar{N}'\xi^1-\bar{N}N_r\left(\xi^0\right)'.
\end{equation}

On the other hand, gauge transformations generated by the constraints can be generally written as 

\begin{equation}
\delta_\varepsilon F=\{F,H_T^{eff}[\varepsilon^0,\varepsilon^1]\}, \quad \delta_\varepsilon N^A=\dot{\varepsilon}^A+N^B\varepsilon^DF^A_{BD}.
\label{deltaFN}
\end{equation} 
Here, $F$ is an arbitrary phase space function and $N^A=(N,N_r)$ are the two Lagrange multipliers. Also $H_T^{eff}[\varepsilon^0,\varepsilon^1]=H_{eff}[\varepsilon^0]+C_c[\varepsilon^1]$, where $\varepsilon^A$ are the parameters of the transformations. Finally, $F^A_{BD}$ are the structure functions of the algebra of constraints $\{C_A,C_B\}=F^D_{AB}C_D$ with $C_A=(H_{eff},C_c)$. In particular, for the case of a generic correction function $\beta(\bar{K}_\varphi)$, we have that

\begin{equation}
\delta_\varepsilon\beta(\bar{K}_\varphi)=\frac{d\beta}{d\bar{K}_\varphi}\left[\dot{\bar{K}}_\varphi\xi^0+\bar{K}'_\varphi\xi^1+\gamma N\frac{\sqrt{\left|E^r\right|}}{(E^\varphi)^2}\left(E^r\right)'\left(\xi^0\right)'\right].
\label{deltaAG}
\end{equation} 

To ease the comparison between the two types of transformations (gauge and coordinates), the past expressions are written in terms of the infinitesimal vector $\xi^\mu$ that generates the diffeomorphism, instead of the parameters of the gauge transformation $\varepsilon^A$. The relations between both of them are $\varepsilon^0=N\xi^0$ and $\varepsilon^1=\xi^1+N_r\xi^0$. From (\ref{deltaAG}) it is also clear that holonomy modifications of the form $\beta(\bar{K}_\varphi)$ do not transform as scalars due to the last term in $\delta_\varepsilon\beta$. This allows us to write

\begin{eqnarray}
\delta_\varepsilon\bar{q}_{rr}&=&\dot{\bar{q}}_{rr}\xi^0+\bar{q}'_{rr}\xi^1+2\bar{q}_{rr}\left[N_r\left(\xi^0\right)'+\left(\xi^1\right)'\right]+\gamma N\frac{d}{d\bar{K}_\varphi}\left(\frac{1}{d\beta_2/d\bar{K}_\varphi}\right)\frac{\left(E^r\right)'}{\sqrt{\left|E^r\right|}}\left(\xi^0\right)' \nonumber \\
&=&\delta_\xi\bar{q}_{rr}+\gamma N\alpha_2^2\frac{d}{d\bar{K}_\varphi}\left(\frac{1}{d\gamma_2/d\bar{K}_\varphi}\right)\frac{\left(E^r\right)'}{\sqrt{\left|E^r\right|}}\left(\xi^0\right)',
\label{dqbrr}
\end{eqnarray}
for $\bar{q}_{rr}=(E^\varphi)^2/(E^rd\beta_2/d\bar{K}_\varphi)$. It is clear from the second line of (\ref{dqbrr}) that, because an additional term appears, the gauge transformation $\delta_\varepsilon\bar{q}_{rr}$ generated by the constraints does not coincide with the usual transformation under diffeomorphisms $\delta_\xi\bar{q}_{rr}$. Thus, an emergent metric formed by the component $\bar{q}_{rr}$ is not covariant. The responsible of this breakdown of covariance can be directly tracked back to the transformation law of holonomy corrections. They do not transform as scalars. For a modified lapse with an arbitrary holonomy corrected factor, for instance, $$\bar{N}=\beta(\bar{K}_\varphi)N,$$ one should expect a similar result. Indeed,

\begin{eqnarray}
\delta_\varepsilon\bar{N}&=&\left(\bar{N}\xi^0\right)\dot{}+\bar{N}'\xi^1-\bar{N}N_r\left(\xi^0\right)'+\gamma N^2\frac{d\beta}{d\bar{K}_\varphi}\frac{\sqrt{\left|E^r\right|}}{(E^\varphi)^2}\left(E^r\right)'\left(\xi^0\right)' \nonumber \\
&=&\delta_\xi\bar{N}+\gamma N^2\frac{d\beta}{d\bar{K}_\varphi}\frac{\sqrt{\left|E^r\right|}}{(E^\varphi)^2}\left(E^r\right)'\left(\xi^0\right)'.
\label{dNb}
\end{eqnarray}
Once again, the manner in which holonomy modifications transform impede covariance. Inverse triad corrections, though, do not represent any obstacle for covariant metrics in this case. This is due to the absence of the $\alpha_1(E^r)$ function in all of the above calculations despite the fact of using the effective Hamiltonian constraint $H_{eff}[N]$ containing said modification. For this reason, from this point forward, we will be forced to treat separately each of the effective contributions of these corrections. The way in which the modified solutions of section V are expressed lets us easily identify them.

Before moving on, we should mention that an alternative proposal, based on the divergence of an effective Einstein tensor, was recently put forth \cite{DelAguila_2025}. In this reference we erroneously concluded that the holonomy corrections given by the functions $\beta_{1,2}$ can lead to covariant models, even in the case of a phase-space dependent holonomy parameter. The mistake in the analysis was to treat $\beta_2$ as a scalar quantity, evidently yielding wrong results\footnote{We thank Martin Bojowald for bringing this point to our attention.}. Regardless, conclusions for the inverse triad corrections of the reference do still hold and are consistent with the preceding arguments.

\subsection{VI.A. Effective Space-Time with Inverse Triad Corrections}

As previously explained, the inclusion of inverse triad corrections in the emergent metric scheme does not lead to the loss of covariance. One can therefore interpret the past effective solutions as coefficients of a modified metric written in different coordinates, i.e., different gauges are connected by coordinate transformations. Ignoring holonomy effects, this is, setting $\delta_\varphi=0$, the exterior and interior metrics, as well as that in PG coordinates are respectively,

\begin{eqnarray}
ds^2&=&\bar{\delta}_r^2\left[-\frac{e^{2R(\bar{r})}}{\bar{r}^2}\left(1-2\bar{m}e^{-R(\bar{r})}\right)d\bar{t}^2+\frac{d\bar{r}^2}{1-2\bar{m}e^{-R(\bar{r})}}+\bar{r}^2d\Omega^2\right], \quad e^{R(\bar{r})}\geq2\bar{m}, \label {IText} \\
ds^2&=&\bar{\delta}_r^2\left[-\frac{d\bar{t}^2}{2\bar{m}e^{-T(\bar{t})}-1}+\frac{e^{2T(\bar{t})}}{\bar{t}^2}\left(2\bar{m}e^{-T(\bar{t})}-1\right)d\bar{r}^2+\bar{t}^2d\Omega^2\right], \quad 0\leq e^{T(\bar{t})}\leq2\bar{m}, \label {ITint} \\
ds^2&=&\bar{\delta}_r^2\left[-\frac{e^{2R(\bar{r})}}{\bar{r}^2}d\bar{t}_{PG}^2+\left(d\bar{r}+\frac{\sqrt{2\bar{m}e^{R(\bar{r})}}}{r}d\bar{t}_{PG}\right)^2+\bar{r}^2d\Omega^2\right], \quad e^{R(\bar{r})}\geq0.
\label {ITPG}
\end{eqnarray}
The previously introduced scaled coordinates, denoted by a bar, are used to express these metrics. It can be easily verified that the exterior and the PG line elements in equations (\ref{IText}) and (\ref{ITPG}) are related by the change of coordinates implied by $$d\bar{t}=d\bar{t}_{PG}-\frac{\bar{r}\sqrt{2\bar{m}e^{-R(\bar{r})}}d\bar{r}}{e^{R(\bar{r})}\left(1-2\bar{m}e^{-R(\bar{r})}\right)}.$$ Meanwhile, the exterior and the interior metrics can be obtained one from another by interchanging $\bar{t}\leftrightarrow\bar{r}$ and reordering the negative signs. Since the metrics were constructed using solutions of the effective dynamics in different gauges, these constitute explicit proofs of the diffeomorphism invariance within the effective theory. Now we can proceed to make meaningful physical and geometric predictions without worrying that such results will depend on the coordinate system chosen to describe them.

\begin{figure}[h]
	\centering
		\includegraphics[scale=0.4]{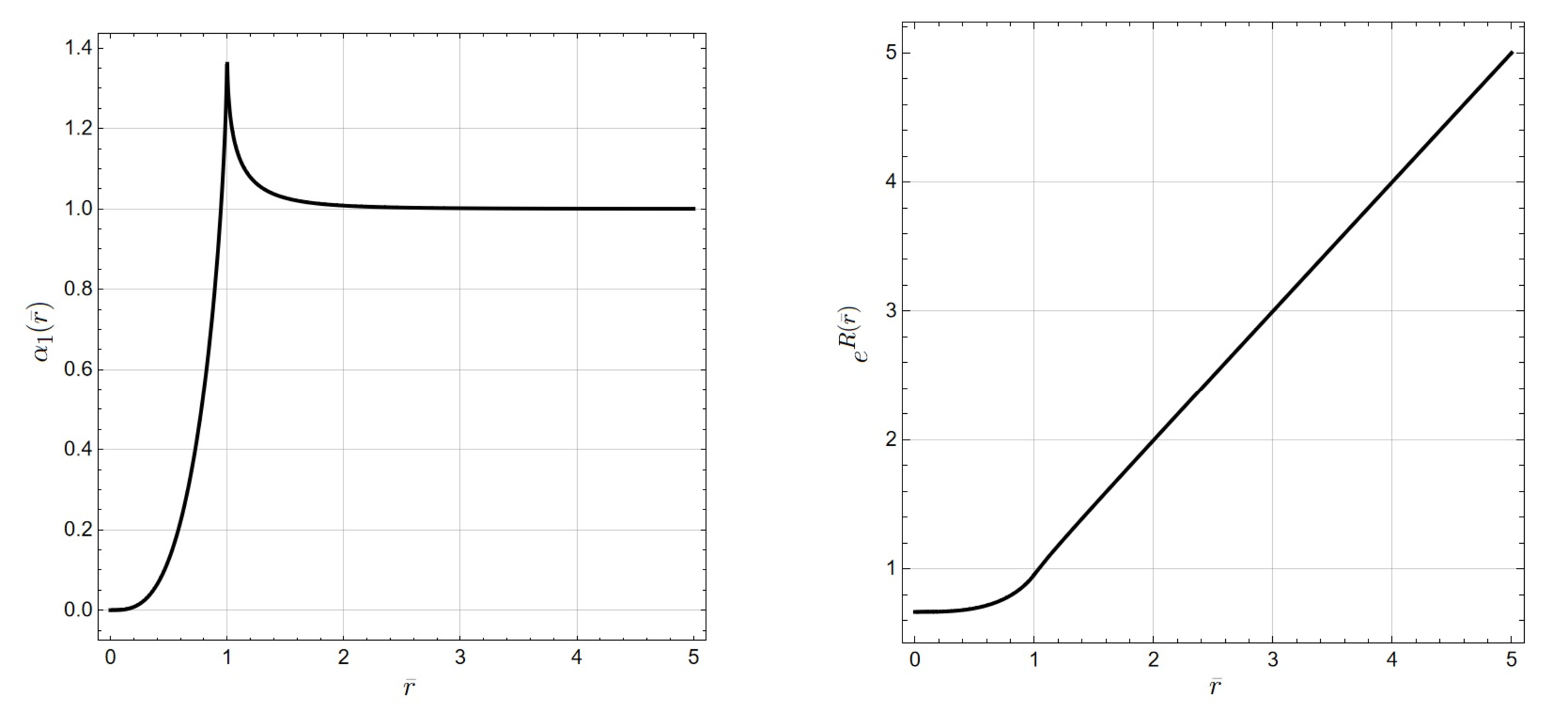}
	\caption{The functions $\alpha_1(\bar{r})$ (left) and $e^{R(\bar{r})}$ (right) as defined by equations (\ref{alpha1}) and (\ref{eRr}), respectively. It can be appreciated that the minimal value of the latter function is $e^{R(0)}=e^{(\pi/2-1)/2}\approx0.665\neq0$.}
	\label{fig:eR(r)}
\end{figure}

The main interest for effective space-times of this type is to compare them with the Schwarzschild geometry and find out to what extent quantum corrections modify its well-known properties. Let us first analyze the event horizon of the black hole with inverse triad modifications. As in the classical case, the canonical vector for coordinate time $\chi=\partial/\partial\bar{t}$ is a Killing vector in the exterior metric (\ref{IText}), and the condition $\chi_\mu\chi^\mu=0$ (the vanishing of its norm) defines the Killing horizon that corresponds to the event horizon. Its radius $\bar{r}_H$ is thus $e^{R(\bar{r}_H)}=2\bar{m}$. It can be shown that $e^{R(\bar{r})}$ is an increasingly monotonic function that behaves linearly for large $\bar{r}$, which is of course the correct classical limit (see figure \ref{fig:eR(r)}).

Due to the previous properties, $e^{R(\bar{r}_H)}=2\bar{m}$ has a single solution and the effective black hole has only one horizon. Note that, as an evident consequence of covariance, one can arrive at this conclusion regardless of which form of the line elements (\ref{IText}-\ref{ITPG}) is chosen for the analysis, the result does not depend on the coordinate system. Therefore, quantum effects do not add another horizon to the usual classical horizon in the Schwarzschild geometry. Quantitatively there is also a minor difference for the possible numerical values of the effective radius $r_H^{eff}$ and the classical radius $r_H^{(cl)}=2m$. For $\bar{m}=1$ for example, $\bar{r}_H^{(cl)}=2$ and $\bar{r}_H^{(eff)}\approx2.004$. This difference becomes even smaller for larger values of the scaled mass $\bar{m}$.

Next, we will examine the interior region of the black hole in search for possible curvature singularities, these are, unbounded curvature scalars such as the Ricci scalar $\mathcal{R}=R^\mu_{\ \mu}$ or Kretschmann scalar $\mathcal{K}=R_{\alpha\beta\mu\nu}R^{\alpha\beta\mu\nu}$. A straightforward calculation yields

\begin{align}
\mathcal{R}=\frac{2e^{-R(\bar{r})}}{\bar{\delta}_r\bar{r}^2}&\left[\left(2\bar{m}+\alpha_1e^{R(\bar{r})}\right)\left(1-\alpha_1\right)-\left(e^{R(\bar{r})}-\bar{m}\right)\bar{r}\frac{d\alpha_1}{d\bar{r}}\right], \nonumber \\
\mathcal{K}=\frac{4e^{-2R(\bar{r})}}{\bar{\delta}_r^4\bar{r}^4}&\left(2\bar{m}^2(2+\alpha_1^2)+2\left[e^{R(\bar{r})}-2\bar{m}+\left(\bar{m}-e^{R(\bar{r})}\right)\alpha_1\right]^2\right. \nonumber \\
&\left.+\left[2e^{R(\bar{r})}-4\bar{m}+\left(2\bar{m}-3e^{R(\bar{r})}\right)\alpha_1+e^{R(\bar{r})}\alpha_1^2+\left(e^{R(\bar{r})}-\bar{m}\right)\bar{r}\frac{d\alpha_1}{d\bar{r}}\right]^2\right).
\label{RK}
\end{align}
It can be readily seen that in the classical case, $\alpha_1=1$, the Ricci scalar vanishes and $\mathcal{K}=48m^2/r^6$ as expected. Unfortunately, these scalars are both unbounded, $$\lim_{\bar{r}\rightarrow0}\mathcal{R},\mathcal{K}=\pm\infty.$$ There is hence a curvature singularity when $\bar{r}\rightarrow0$ just as in a classical black hole. The exponential $e^{R(\bar{r})}$ associated to the inverse triad corrections is not responsible for the appearance of said singularity since, as seen in figure \ref{fig:eR(r)}, it is bounded from below. Nevertheless, singular $r^{-n}$ factors remain present in the expressions of (\ref{RK}). Note though that for the Kretschmann scalar $\mathcal{K}^{(eff)}\sim r^{-4}$ in the effective case, while $\mathcal{K}^{(cl)}\sim r^{-6}$ classically. This can be interpreted as quantum effects due to inverse triads decreasing the strength of the singularity, despite them not being enough to remove it completely. 

It can be concluded then that, on their own, inverse triad corrections in LQG do not lead to significant qualitative changes in the effective geometry compared to its classical counterpart. Both space-times contain an event horizon, and inside it, a curvature singularity. Holonomy modifications are still missing, which may lead to more interesting results regarding the boundedness of curvature scalars.

Let us consider, though, geodesics approaching the ill-defined region of space-time at $\bar{r}=0$. The question of what consequences will the curvature singularity have on the completeness of these curves is particularly relevant. This comes from the fact that geodesic completeness has been used as a sufficient criteria to characterize a space-time as singular. In the classical Schwarzschild case, the presence of divergent curvature leads to null and time-like geodesic incompleteness. In the following we analyze if loop quantum effects have any impact on this specific property of the effective space-time.

For the study of geodesics it is standard to focus on conserved quantities along the paths of such curves. Since $\chi=\partial/\partial\bar{t}$ and $\zeta=\partial/\partial\varphi$ are Killing vectors, the quantities $p_{\bar{t}}=\chi_\mu v^\mu$ and $p_\varphi=\zeta_\mu v^\mu$ associated to them are constants of motion. Here, $v^\mu=dx^\mu/d\lambda$ is the coordinate velocity with respect to an affine parameter $\lambda$. Explicitly we have that, 

\begin{equation}
p_{\bar{t}}=-\frac{\bar{\delta_r}^2}{\bar{r}^2}e^{2R(\bar{r})}\left(1-2\bar{m}e^{-R(\bar{r})}\right)\frac{d\bar{t}}{d\lambda}, \quad p_\varphi=\bar{\delta}_r^2\bar{r}^2\sin^2\theta\frac{d\varphi}{d\lambda}.
\label{p}
\end{equation}
The past quantities are commonly given the physical interpretation at asymptotic infinity of the energy $E=-p_{\bar{t}}/\bar{\delta}_r$ and angular momentum (with respect to the $z$-axis) $l=p_\varphi$ of a freely-falling test particle. An additional conserved quantity is the norm of the four-velocity $v^\mu$, namely,

\begin{equation}
v_\mu v^\mu=\kappa=\begin{cases} 0, & \text{for null geodesics}, \\
    -1, & \text{for time-like geodesics}.
\end{cases}
\label{kappa}
\end{equation}
If we consider only radial geodesics, i.e., $d\theta/d\lambda=d\varphi/d\lambda=0$, then (\ref{p}) and (\ref{kappa}) can be combined to obtain the following expression,

\begin{equation}
\left(\frac{d\bar{r}}{d\bar{\lambda}}\right)^2=\frac{E^2\bar{r}^2}{e^{2R(\bar{r})}}+\kappa V(\bar{r}),
\label{vr}
\end{equation} 
with a potential given by $V(\bar{r})=1-2\bar{m}e^{-R(\bar{r})}$ and a rescaled affine parameter $\bar{\lambda}=\lambda/\bar{\delta}_r$. From the above equation the affine parameter $\lambda$ can be written as an integral

\begin{equation}
\bar{\lambda}-\bar{\lambda_0}=\pm\int\frac{d\bar{r}}{\sqrt{E^2\bar{r}^2e^{-2R(\bar{r})}+\kappa V(\bar{r})}},
\label{lambda}
\end{equation}
where the $+(-)$ sign describes outgoing (ingoing) geodesics, and $\bar{\lambda}_0$ is an integration constant. Furthermore, for radial geodesics with constant $E$ the only non-trivial component of the geodesic equation $v^\mu\nabla_\mu v^\nu=0$ is the radial one, which in terms of constants of motion reads 

\begin{equation}
\frac{d^2\bar{r}}{d\bar{\lambda}^2}=E^2\bar{r}e^{-2R(\bar{r})}\left(1-\alpha_1\right)+\kappa\alpha_1e^{-R(\bar{r})}\frac{\bar{m}}{\bar{r}}.
\label{ar}
\end{equation}

Even though an analytical expression for $\bar{r}(\bar{\lambda})$ cannot be found from the previous relations, relevant information can still be extracted. Let us start with null geodesics ($\kappa=0$), the affine parameter and radial acceleration reduce respectively to $$\Lambda=\pm\int\frac{e^{R(\bar{r})}}{\bar{r}}d\bar{r}, \quad \frac{d^2\bar{r}}{d\bar{\lambda}^2}=E^2\bar{r}e^{-2R(\bar{r})}\left(1-\alpha_1\right),$$ where we defined $\Lambda=E(\bar{\lambda}-\bar{\lambda}_0)$. Note that the integrand for $\Lambda$ diverges as $\bar{r}\rightarrow0$ and hence, so does the integral. To see this, consider a series expansion around $\bar{r}=0$ for the exponential $e^{R(\bar{r})}$ of the numerator. Such expansion always exists because the function is analytical in that point. Then $\Lambda=a_0\ln|\bar{r}/\bar{r}_0|+\mathcal{O}(\bar{r})$ for constant $a_0$ and $\bar{r}_0$, thus $\Lambda\rightarrow\pm\infty$ as $\bar{r}\rightarrow0$. Put into words, an infinite amount of affine parameter is needed to reach $\bar{r}=0$. This result is consistent with the expressions for radial velocity and acceleration of (\ref{vr}) and (\ref{ar}). For radial null geodesics, $$\left.\frac{d\bar{r}}{d\bar{\lambda}}\right|_{\bar{r}=0}=\left.\frac{d^2\bar{r}}{d\bar{\lambda}^2}\right|_{\bar{r}=0}=0,$$ and the curves slow down when approaching the singularity, needing increasingly more affine parameter $\bar{\lambda}$ to keep advancing toward it. It can be concluded hence that the curvature singularity does not cause null geodesic incompleteness for purely radial curves. This is not true for the Schwarzschild space-time for which $e^{R(\bar{r})}=\bar{r}$ and $\alpha_1=1$. In that case radial null geodesics can be easily integrated, yielding $\bar{r}=\pm\Lambda$, and therefore the singularity can be reached in a finite affine parameter. These are incomplete curves because $\bar{r}=0$ is an ill-defined space-time region. It can be seen that these contrasting geodesic behaviors in the classical and effective metrics arise as a result of including inverse triad corrections contained in the exponential $e^{R(\bar{r})}$. We have described a subtle, but drastic, difference between the effective and classical black holes due to loop quantum modifications.

We turn now to time-like geodesics ($\kappa=-1$). For these curves the potential $V(\bar{r})=1-2\bar{m}e^{-R(\bar{r})}$ becomes relevant. Recall that inside the horizon we have that $e^{R(\bar{r})}<2\bar{m}$, which implies that $\kappa V(0)>0$, and the integrand in (\ref{lambda}) is well-defined for $\bar{r}=0$. This means that one can always find a unique solution for the initial value problem defined by the differential equation (\ref{vr}) and the condition $\bar{r}(\bar{\lambda}_0)=0$ for finite $\bar{\lambda}_0$. These curves are thus incomplete, as they reach the singular region in a finite affine parameter. Examining the radial velocity (\ref{vr}) and radial acceleration (\ref{ar}), it can be seen that $$\left.\left(\frac{d\bar{r}}{d\bar{\lambda}}\right)^2\right|_{\bar{r}=0}>0, \quad \lim_{\bar{r}\rightarrow0}\frac{d^2\bar{r}}{d\bar{\lambda}^2}=-\infty.$$ This is physically consistent as well. Since the velocity at $\bar{r}=0$ is non-vanishing and the acceleration near that region is negative, time-like observers in free-fall feel an attractive potential toward the singularity, just as in the case of the Schwarzschild black hole. Unfortunately, we can realize then that time-like geodesics in the effective model follow a similar behavior as those of the classical theory, they are indeed incomplete due to the singularity.

This effective geometry modified by inverse triad corrections is a peculiar example of a space-time in which a curvature singularity does imply time-like geodesic incompleteness, but not of the null type. In this sense, it could be said that loop quantum effects restore completeness for null geodesics, but fail to do so for those that are time-like. The main reason behind this being that the attractive potential, which affects only time-like observers, suffers little qualitative modifications with respect to the classical case. One can argue, though, that the effects of the attractive potential are reduced in the modified model. In the Schwarzschild metric the potential is $V_{cl}(r)=1-2m/r$ and thus $\lim_{r\rightarrow0}V_{cl}(r)=-\infty$, whereas in the inverse triad corrected geometry we have that $-\infty<V(0)<0$ is still negative but finite. Nevertheless, this boundedness property is not enough to recover time-like geodesic completeness.

To give additional information and provide further support to the past claims, the expansion $\Theta$ of the congruence formed by the previously described geodesics can be computed. From the coordinate velocities calculated in (\ref{p}) and (\ref{vr}), the tangent to these curves can be expressed in the scaled coordinate basis $\{\bar{t},\bar{r},\theta,\varphi\}$ as $$v_\pm^\mu=\left(\frac{E\bar{r}^2}{e^{2R(\bar{r})}(1-2\bar{m}e^{-R(\bar{r})})},\pm\sqrt{\frac{E^2\bar{r}^2}{e^{2R(\bar{r})}}+\kappa V(\bar{r})},0,0\right).$$ For the congruence of ingoing geodesics the expansion becomes $\Theta_-=\nabla_\mu v_-^\mu$, which turns into 

\begin{equation}
\Theta_-=-\frac{2}{\bar{r}}\sqrt{\frac{E^2\bar{r}^2}{e^{2R(\bar{r})}}+\kappa V(\bar{r})}.
\end{equation} 
It is clear that the time-like expansion ($\kappa=-1$) will diverge in the limit, $\lim_{\bar{r}\rightarrow0}\Theta_-=-\infty$. This means that time-like geodesics encounter a focal point which corresponds to the singularity at $\bar{r}=0$. On the other hand, if $\kappa=0$, then the null expansion reduces to 

\begin{equation}
\Theta_-=-2Ee^{-R(\bar{r})}.
\label{effexp}
\end{equation}
Since $e^{R(\bar{r})}$ has a non-zero minimal value (see figure \ref{fig:eR(r)}), this quantity is bounded for finite energy, which implies that null geodesics will keep focusing toward $\bar{r}=0$ without ever reaching the focal point. This is consistent with the fact that the singularity does not cause null geodesics incompleteness in the effective case. By contrast, in the classical theory $e^{R(\bar{r})}=\bar{r}$ and thus, $\Theta_-^{(cl)}$ does diverge when $\bar{r}\rightarrow0$, leading to incomplete null curves.

It is well-known that if a congruence of geodesics is initially converging ($\Theta_0<0$) and the energy conditions hold, then by the Raychaudhuri equation its expansion will be unbounded \cite{wald}. The effective null expansion (\ref{effexp}), however, is negative everywhere but bounded nevertheless. Energy conditions must therefore be violated somewhere in the quantum-corrected space-time. Indeed, a straightforward calculation yields $$R_{vv}^{(-)}=R_{\mu\nu}v_-^{\mu}v_-^{\nu}=2E^2e^{-R(\bar{r})}(\alpha_1-1),$$ for null geodesics. From figure \ref{fig:eR(r)}, one can realize that $\alpha_1-1<0$ for $0<\bar{r}<\bar{r}_0\approx1$, and thus $R_{vv}^{(-)}$ can be negative. The null energy condition, which states that $R_{\mu\nu}k^{\mu}k^{\nu}\geq0$ everywhere for all null $k^\mu$, fails to be satisfied. Furthermore, the violation takes place in what can be considered the Planck regime, namely, in values $\bar{r}<1$ (recall that $\bar{r}=r/\bar{\delta}_r$ and $\bar{\delta}_r\sim l_{Pl}$). The breakdown of the null energy condition in semiclassical spherical metrics was also reported in \cite{Gambini2020}. We can return once again to the Schwarzschild space-time by setting $\alpha_1=1$, and recover $R_{vv}^{(-)}=0$ as expected, satisfying consequently the null energy condition.

In the celebrated singularity theorem by Penrose, one of the sufficient conditions for the existence of incomplete null geodesics is the fulfillment of the null energy condition combined with the presence of closed trapped surfaces and a non-compact Cauchy surface \cite{penrose}. In this effective geometry, the violation of said energy condition is then crucial for avoiding null geodesic incompleteness, and is the reason why the theorem does not hold for this case. The quantum effects described by inverse triad corrections can be interpreted to be responsible for this unusual behavior.

Finally, we should be cautious with the differentiability of the correction function $\alpha_1$ due the consequences it may have, particularly in the curvature of the effective geometry. From figure \ref{fig:eR(r)} it is readily seen that $\alpha_1(\bar{r})$ is not differentiable at $\bar{r}=1$. This can be analytically reinforced by examining equation (\ref{alpha1}). Since the Ricci and Kretschmann scalars in (\ref{RK}) contain derivatives $d\alpha_1/d\bar{r}$, the lack of differentiability becomes problematic because it then leads to ill-defined curvature at $\bar{r}=1$, thus adding another curvature singularity to the effective metric. We will postpone further discussion of the interpretation and meaning of this singularity, as well as possible ways to solve this issue, for future work.

\subsection{VI.B. Holonomy Solutions and Their Possible Regularity}

While inverse triad corrections on their own yield covariant solutions that can be readily interpreted as the components of an effective metric, the holonomy modifications found in this paper do not possess this property, as explained at the start of this section. In this subsection we describe preliminary results that would allow us to give certain physical meaning to this type of effective solutions. The main focus will be of determining whether they describe a singular or regular theory in a suitable sense.

In GR the metric is a mathematical tool that, combined with other tensors, allows the calculation of meaningful physical or geometrical scalars. Predictions are then often made based on these quantities. For instance, the vanishing of the norm of a time-like Killing vector field defines a Killing horizon, information about the space-time curvature can be extracted from the Ricci scalar, the energy measured by an observer with tangent $v^\mu$ is given by $T_{\mu\nu}v^\mu v^\nu$, etc. All of these are scalars. In what follows we will put forth a quantity that changes as a scalar under gauge transformations in the effective theory and that may be relevant for the singularity evasion question.

Consider the initial spherically symmetric line element (\ref{ds2Esf}) in a homogeneous gauge (only time dependence). The standard analysis utilized in the last subsection to obtain geodesics through conserved quantities can be applied to this metric, yielding the following radial null geodesics $$k_-^\mu=\left(-\frac{\sqrt{|E^r|}}{NE^\varphi},\frac{E^r}{\left(E^\varphi\right)^2}\left[1-\frac{N_rE^\varphi}{N\sqrt{|E^r|}}\right],0,0\right).$$ The unscaled coordinate basis is now being used $\{t,r,\theta,\varphi\}$ to write the vector $k_\pm^\mu$. The expansion of the congruence defined by $k_-^\mu$ is given by 

\begin{equation}
\Theta_-=-\frac{\dot{E}^r}{NE^\varphi\sqrt{|E^r|}}.
\label{exphom}
\end{equation}
This scalar was of course obtained by the usual methods of the classical theory in which covariance is an underlying property of the system. This property is absent in the effective model. Let us postulate regardless, that $\Theta_-$ as expressed by (\ref{exphom}) is a relevant quantity in the semiclassical theory, and that such quantity is still related to the expansion of radial null geodesics guided by classical arguments.

With the aid of the Dirac observable, which in the case of discarding inverse triad corrections and keeping only holonomy modifications turns into $\mathcal{O}_{eff}=m$, the effective field equations can be solved without fixing the gauge. This observable is particularly helpful because it gives an algebraic relation of the phase space variables. To avoid repetition, the details of this process will be omitted here, but are analogous to those outlined in the corresponding subsections of section V. The final result is

\begin{equation}
N^2=\frac{\left(\dot{E}^r\right)^2}{4E^r\left[2m/\sqrt{|E^r|}-1\right]\left[1-\gamma^2\delta_\varphi^2\left(2m/\sqrt{|E^r|}-1\right)/4\right]}, \quad \left(E^\varphi\right)^2=\frac{1}{4}\left(\frac{\dot{E}^r}{N}\right)^2, \quad \beta_2=\frac{\gamma\dot{E}^r}{2N\sqrt{|E^r|}}.
\label{homholsol}
\end{equation}
The solution is written in a manner in which gauge freedom can be explicitly seen. As stated previously, there are two degrees of freedom related to gauge. Here, they manifest as the liberty of choosing $E^r$ and $N_r$ to express the rest of the canonical variables and lapse function. The radial shift $N_r$ does not appear in the solution, though, because it decouples from the effective field equations in the homogeneous gauge, see (\ref{EFE}). Indeed, it can be verified that if the gauge is fixed to $E^r=t^2$, then the past equations coincide with the interior solution of subsection V.B. when further making $\alpha_1=1$.

Inserting (\ref{homholsol}) in the expansion then yields $$\Theta_-=-\frac{2}{\sqrt{|E^r|}},$$ which can be shown to be bounded. Recall that $\beta_2=\sin(\delta_\varphi\bar{K}_\varphi)/\delta_\varphi$ and hence, $0\leq\delta_\varphi^2\beta_2^2\leq1$. If the effective homogeneous solution is considered this inequality becomes, 

\begin{equation}
0\leq\gamma^2\delta_\varphi^2\left[\frac{2m}{\sqrt{|E^r|}}-1\right]\left[1-\frac{\gamma^2\delta_\varphi^2}{4}\left(\frac{2m}{\sqrt{|E^r|}}-1\right)\right]\leq1.
\label{homineq}
\end{equation}
The only way that the lower bound of inequality (\ref{homineq}) can be satisfied is if $$\frac{2\gamma^2\delta_\varphi^2m}{4+\gamma^2\delta_\varphi^2}\leq\sqrt{|E^r|}\leq2m.$$ Note that this is simply the restricted domain (\ref{intdom}) of the interior solution without inverse triad effects, but generalized to an arbitrary homogeneous gauge. The lower bound of the past inequality has also been found in \cite{asier} within the context of covariant spherically symmetric effective models\footnote{The holonomy parameter used in this paper differs by a numerical factor from that of \cite{asier}.}. Therefore, the congruence of radial null geodesics has a bounded expansion for all homogeneous gauges.

We proceed now to examine the inhomogeneous gauge following a similar analysis as before. The tangent vector to ingoing radial null geodesics in the spherical metric (\ref{ds2Esf}) without time dependence is $$k_-^\mu=\left(\frac{1}{N\left(N+E^\varphi N_r/\sqrt{|E^r|}\right)},-\frac{\sqrt{|E^r|}}{NE^\varphi},0,0\right),$$ where again the coordinate basis $\{t,r,\theta,\varphi\}$ is utilized. The corresponding expansion of these curves reads $$\Theta_-=-\frac{\left(E^r\right)'}{NE^\varphi\sqrt{|E^r|}}.$$ The Dirac observable $\mathcal{O}_{eff}=m$ is used to find the most general inhomogeneous solution without additional gauge fixing. Once more, the effective solution will be written in a way that reflects its dependency on the gauge specified by $E^r$ and $E^\varphi$, namely,

\begin{align}
N_r^2=&\frac{E^r}{\left(E^\varphi\right)^2}\left[\frac{2m}{\sqrt{|E^r|}}-1+\left(\frac{(E^r)'}{2E^\varphi}\right)^2\right]\left[1-\frac{\gamma^2\delta_\varphi^2}{4}\left(\frac{2m}{\sqrt{|E^r|}}-1+\left(\frac{(E^r)'}{2E^\varphi}\right)^2\right)\right], \nonumber \\
N=&\frac{(E^r)'}{2E^\varphi}, \quad \beta_2=-\frac{\gamma\delta_\varphi N_r\left(E^r\right)'}{2N\sqrt{|E^r|}}. 
\end{align}
The exterior solution of the past section (without inverse triad corrections) can be recovered from this general form if one chooses $E^r=r^2$ and $\left(E^\varphi\right)^2/E^r=1+\gamma_n^2-2m/r$. Likewise, the holonomy modified PG solution can be found by setting $E^r=\left(E^\varphi\right)=r^2$. From these expressions it is clear that the expansion reduces to $$\Theta_-=-\frac{2}{\sqrt{|E^r|}},$$ i.e., the same phase space function as in the homogeneous gauge. It can also be proved to be bounded due to holonomy corrections. The argument goes along the same lines as the previous one. For this effective solution, the inequality $0\leq\delta_\varphi^2\beta_2^2\leq1$ turns into 

\begin{equation}
0\leq\gamma^2\delta_\varphi^2\mathcal{E}\left(1-\frac{1}{4}\gamma^2\delta_\varphi^2\mathcal{E}\right)\leq1,
\label{inhomineq}
\end{equation} 
with $\mathcal{E}=2m/\sqrt{|E^r|}-1+N^2$. The lower bound of the previous inequality can only be satisfied if $0\leq\mathcal{E}\leq4/\gamma^2\delta_\varphi^2$. However, $\lim_{\sqrt{|E^r|}\rightarrow0}\mathcal{E}=\infty$, which contradicts (\ref{inhomineq}). Thus, $\sqrt{|E^r|}$ cannot be arbitrarily small and $\Theta_-$ does not grow without bound.

Let us summarize these results. The quantity $\Theta_-=-2/\sqrt{|E^r|}$ can be given the meaning of the expansion of radial null geodescis in the effective theory. It coincides in both homogeneous and inhomogeneous gauges, $\Theta_-^{(hom)}=\Theta_-^{(inhom)}=\Theta_-$, and is bounded due to holonomy effects. In contrast by taking the classical limit, $\delta_\varphi\rightarrow0$, the expansion is unrestricted and can diverge. Furthermore, $\Theta_-$ is a scalar quantity. This follows from the fact that any phase space function $\alpha(E^r)$ transforms as a scalar under gauge transformations generated by the constraints, i.e., $$\delta_\varepsilon\alpha(E^r)=\frac{d\alpha}{dE^r}\left[\dot{E}^r\xi^0+\left(E^r\right)'\xi^1\right].$$ Due to the above reasons, we argue that $\Theta_-$ is a meaningful quantity within the holonomy-corrected model studied in this paper. Holonomy effects restrict the domain of the homogeneous and inhomogeneous solutions found here, which then leads to the boundedness of the expansion. We interpret this as a manifestation that singularities may be absent from the model. Further evidence in this direction is needed in order for this last claim to be robust.

Before ending this section it is important to highlight the following. The construction and consideration of the expansion $\Theta_-$ as a relevant quantity in the effective model can be deceptive. Its explicit form was extracted from a classical analysis and then reduced to a simpler expression using the effective solutions in different gauges, because of this, it may seem that one can apply the same steps to any space-time scalar and then propose it as a quantity of interest. It turns out that this cannot be done consistently due to the breakdown of covariance. For example, if one considers the expansion $\Theta_{TL}$ of radial time-like geodesics, issues begin to appear in the analysis. To start with, $\Theta_{TL}$ seen as an on-shell phase space function, is not the same in the homogeneous and inhomogeneous gauges, this is, $\Theta_{TL}^{(hom)}(E^r)\neq\Theta_{TL}^{(inhom)}(E^r,E^\varphi)$, where the explicit equations are not shown. In addition, since $\Theta_{TL}^{(inhom)}$ depends on $E^\varphi$, it does not transform as a scalar. The expansion $\Theta_-$ thereby is a special case which happens to reunite all of the desired characteristics. At present we do not know if other such quantities, related in some way to singularities, exist within this effective model.

\section{VII. Conclusions and Discussion}

In this work the path integral method has been used to obtain an effective model that describes quantum corrections in spherically symmetric space-times. Specifically, the semiclassical model derived features quantum effects representing inverse triad and holonomy corrections that generally appear in LQG, modifying hence the Hamiltonian of the classical model. The procedure followed takes into account two holonomy parameters, $\delta_r$ and $\delta_\varphi$. The first one along the radial direction, and the last one along the polar and azimuthal directions. Both parameters are assumed to be constant and independent of the phase space variables. Naturally, these parameters become part of the effective theory, and can be interpreted as reflecting the underlying discrete nature of space-time. Restricting the holonomy parameters to small values reduces the quantum effects and approximates the model to the classical one.

Among the desired properties of the described effective model is the already mentioned fact that it reduces to the classical Hamiltonian constraint when a suitable limit is taken into account. This limit corresponds to regions where classical behavior is expected and the theory is dominated by General Relativity. Some quantum corrections presented here also coincide with those of other works found by different methods, namely, the expectation value method \cite{LTB}.

Despite this, when computing the constraint algebra of the effective Hamiltonian, it was found that it does not possess the structure followed by the classical theory. This has important implications regarding the general covariance of the model, which is absent due to the violation of the expected algebra. A similar problem was already reported in previous examples \cite{Deformed1, Deformed2}. This led us to simplify the complete effective model and discard those corrections that generated anomalous terms in the algebra. Hence in the simplified model, which contains a reduced number of holonomy and inverse triad corrections, the constraints remain first class and still generate gauge transformations. In order to obtain this property at the effective level, we were forced to slightly modify the usual quantization procedure and shrink the loop along the angular directions used to approximate curvature in terms of holonomies. The effective field equations yielded by the simplified model were solved in different gauges and a physical meaning was given to some of these solutions. 

It turned out that the holonomy modifications found by the path integral method are not covariant, even in the simplified model. On the contrary, inverse triad corrections are compatible with a diffeomorphism invariant model. For this reason, we were able to write an explicit effective metric with the latter type of corrections, but not the former. The modified geometry with inverse triad effects does not qualitatively differ from the Schwarzschild black hole for the most part, both feature an event horizon and a curvature singularity inside it. A subtle difference arise, however, when examining the null geodesic completeness in the effective space-time. Radial null geodesics need an infinite affine parameter to reach the singularity and therefore, do not become incomplete due to its presence, as opposed to the classical case in which the same kind of curves end their paths because of it. The effective expansion of this congruence of null geodesics is bounded. The reason behind this contrasting behavior with respect to the Schwarzschild black hole is the violation of the null energy condition in the semiclassical space-time, which occurs in the Planck regime. On the other hand, time-like geodesics are both classically and effectively incomplete, and their expansion unbounded. 

Finally, because the holonomy solutions found here do not transform covariantly, the typical space-time scheme is not available. One is then left with an effective field theory in a phase space with gauge transformations. Aided by classical arguments, we put forth a phase space quantity that transforms as a scalar and can be linked to the expansion of radial null geodesics. Quantum effects restrict the domain of these holonomy solutions, which leads once again to a bounded expansion scalar. This is a positive sign potentially indicating that the effective model is non-singular. Despite this, further work is needed in this direction in order to claim that the classical singularity has been completely resolved. \\


 \textbf{Acknowledgments.} We are grateful to M. Bojowald for discussions on several details of this work. J.C.D.A. acknowledges financial support from SECIHTI postdoctoral fellowships. This work was also partially supported by SECIHTI grant CBF-2023-2024-1937, as well as SNII-786529 (JCDA) and SNII-14585 (HAMT). \\

\section{References}

\bibliography{references}

@article{WDW,
  title = {Quantum Theory of Gravity. I. The Canonical Theory},
  author = {DeWitt, Bryce S.},
  journal = {Phys. Rev.},
  volume = {160},
  issue = {5},
  pages = {1113--1148},
  numpages = {0},
  year = {1967},
  month = {Aug},
  publisher = {American Physical Society},
  doi = {10.1103/PhysRev.160.1113},
  url = {https://link.aps.org/doi/10.1103/PhysRev.160.1113}
}

@article{AshtekarLewandowski,
doi = {10.1088/0264-9381/21/15/R01},
url = {https://dx.doi.org/10.1088/0264-9381/21/15/R01},
year = {2004},
month = {jul},
publisher = {},
volume = {21},
number = {15},
pages = {R53},
author = {Abhay Ashtekar and  Jerzy Lewandowski},
title = {Background independent quantum gravity: a status report},
journal = {Classical and Quantum Gravity}
}

@article{LQC,
    author = "Ashtekar, Abhay and Bojowald, Martin and Lewandowski, Jerzy",
    title = "{Mathematical structure of loop quantum cosmology}",
    eprint = "gr-qc/0304074",
    archivePrefix = "arXiv",
    reportNumber = "CGPG-03-4-4",
    doi = "10.4310/ATMP.2003.v7.n2.a2",
    journal = "Adv. Theor. Math. Phys.",
    volume = "7",
    number = "2",
    pages = "233--268",
    year = "2003"
}

@article{Improved,
  title = {Quantum nature of the big bang: Improved dynamics},
  author = {Ashtekar, Abhay and Pawlowski, Tomasz and Singh, Parampreet},
  journal = {Phys. Rev. D},
  volume = {74},
  issue = {8},
  pages = {084003},
  numpages = {23},
  year = {2006},
  month = {Oct},
  publisher = {American Physical Society},
  doi = {10.1103/PhysRevD.74.084003},
  url = {https://link.aps.org/doi/10.1103/PhysRevD.74.084003}
}

@article{SchwarzschildSingularity,
doi = {10.1088/0264-9381/23/2/008},
url = {https://dx.doi.org/10.1088/0264-9381/23/2/008},
year = {2005},
month = {dec},
publisher = {},
volume = {23},
number = {2},
pages = {391},
author = {Abhay Ashtekar and Martin Bojowald},
title = {Quantum geometry and the Schwarzschild singularity},
journal = {Classical and Quantum Gravity}
}

@article{Modesto,
doi = {10.1088/0264-9381/23/18/006},
url = {https://dx.doi.org/10.1088/0264-9381/23/18/006},
year = {2006},
month = {aug},
publisher = {},
volume = {23},
number = {18},
pages = {5587},
author = {Modesto, Leonardo},
title = {Loop quantum black hole},
journal = {Classical and Quantum Gravity}
}

@article{Revisited,
doi = {10.1088/0264-9381/33/5/055006},
url = {https://dx.doi.org/10.1088/0264-9381/33/5/055006},
year = {2016},
month = {feb},
publisher = {IOP Publishing},
volume = {33},
number = {5},
pages = {055006},
author = {Alejandro Corichi and Parampreet Singh},
title = {Loop quantization of the Schwarzschild interior revisited},
journal = {Classical and Quantum Gravity}
}

@article{AOS,
  title = {Quantum extension of the Kruskal spacetime},
  author = {Ashtekar, Abhay and Olmedo, Javier and Singh, Parampreet},
  journal = {Phys. Rev. D},
  volume = {98},
  issue = {12},
  pages = {126003},
  numpages = {29},
  year = {2018},
  month = {Dec},
  publisher = {American Physical Society},
  doi = {10.1103/PhysRevD.98.126003},
  url = {https://link.aps.org/doi/10.1103/PhysRevD.98.126003}
}

@article{AOS2,
  title = {Space of solutions of the Ashtekar-Olmedo-Singh effective black hole model},
  author = {Elizaga Navascu\'es, Beatriz and Garc\'{\i}a-Quismondo, Alejandro and Mena Marug\'an, Guillermo A.},
  journal = {Phys. Rev. D},
  volume = {106},
  issue = {6},
  pages = {063516},
  numpages = {10},
  year = {2022},
  month = {Sep},
  publisher = {American Physical Society},
  doi = {10.1103/PhysRevD.106.063516},
  url = {https://link.aps.org/doi/10.1103/PhysRevD.106.063516}
}

@article{Gambini2020,
doi = {10.1088/1361-6382/aba842},
url = {https://doi.org/10.1088/1361-6382/aba842},
year = {2020},
month = {sep},
publisher = {IOP Publishing},
volume = {37},
number = {20},
pages = {205012},
author = {Gambini, Rodolfo and Olmedo, Javier and Pullin, Jorge},
title = {Spherically symmetric loop quantum gravity: analysis of improved dynamics},
journal = {Classical and Quantum Gravity}
}

@article{PathIntegralSLQC,
  title = {Path integrals and the WKB approximation in loop quantum cosmology},
  author = {Ashtekar, Abhay and Campiglia, Miguel and Henderson, Adam},
  journal = {Phys. Rev. D},
  volume = {82},
  issue = {12},
  pages = {124043},
  numpages = {12},
  year = {2010},
  month = {Dec},
  publisher = {American Physical Society},
  doi = {10.1103/PhysRevD.82.124043},
  url = {https://link.aps.org/doi/10.1103/PhysRevD.82.124043}
}

@article{Hugo1,
title = {Effective dynamics of the Schwarzschild black hole interior with inverse triad corrections},
journal = {Annals of Physics},
volume = {426},
pages = {168401},
year = {2021},
issn = {0003-4916},
doi = {https://doi.org/10.1016/j.aop.2021.168401},
url = {https://www.sciencedirect.com/science/article/pii/S0003491621000075},
author = {Hugo A. Morales-T\'ecotl and Saeed Rastgoo and Juan C. Ruelas}
}

@article{BianchiI,
doi = {10.1088/0264-9381/30/6/065010},
url = {https://doi.org/10.1088/0264-9381/30/6/065010},
year = {2013},
month = {feb},
publisher = {IOP Publishing},
volume = {30},
number = {6},
pages = {065010},
author = {Liu, Xiao and Huang, Fei and Zhu, Jian-Yang},
title = {Path integral of Bianchi I models in loop quantum cosmology},
journal = {Classical and Quantum Gravity}
}

@article{Corichi,
  title = {Geometric perspective on singularity resolution and uniqueness in loop quantum cosmology},
  author = {Corichi, Alejandro and Singh, Parampreet},
  journal = {Phys. Rev. D},
  volume = {80},
  issue = {4},
  pages = {044024},
  numpages = {10},
  year = {2009},
  month = {Aug},
  publisher = {American Physical Society},
  doi = {10.1103/PhysRevD.80.044024},
  url = {https://link.aps.org/doi/10.1103/PhysRevD.80.044024}
}

@article{Hugo2,
  title = {Effective loop quantum geometry of Schwarzschild interior},
  author = {Cortez, Jer\'onimo and Cuervo, William and Morales-T\'ecotl, Hugo A. and Ruelas, Juan C.},
  journal = {Phys. Rev. D},
  volume = {95},
  issue = {6},
  pages = {064041},
  numpages = {14},
  year = {2017},
  month = {Mar},
  publisher = {American Physical Society},
  doi = {10.1103/PhysRevD.95.064041},
  url = {https://link.aps.org/doi/10.1103/PhysRevD.95.064041}
}

@article{Joe,
doi = {10.1088/0264-9381/32/1/015009},
url = {https://dx.doi.org/10.1088/0264-9381/32/1/015009},
year = {2014},
month = {dec},
publisher = {IOP Publishing},
volume = {32},
number = {1},
pages = {015009},
author = {Anton Joe and Parampreet Singh},
title = {Kantowski Sachs spacetime in loop quantum cosmology: bounds on expansion and shear scalars and the viability of quantization prescriptions},
journal = {Classical and Quantum Gravity}
}

@article{RobustnessBounce,
doi = {10.1088/0264-9381/31/10/105015},
url = {https://dx.doi.org/10.1088/0264-9381/31/10/105015},
year = {2014},
month = {may},
publisher = {IOP Publishing},
volume = {31},
number = {10},
pages = {105015},
author = {Peter Diener and Brajesh Gupt and Parampreet Singh},
title = {Numerical simulations of a loop quantum cosmos: robustness of the quantum bounce and the validity of effective dynamics},
journal = {Classical and Quantum Gravity}
}

@article{Deformed1,
  title = {Deformed general relativity and effective actions from loop quantum gravity},
  author = {Bojowald, Martin and Paily, George M.},
  journal = {Phys. Rev. D},
  volume = {86},
  issue = {10},
  pages = {104018},
  numpages = {24},
  year = {2012},
  month = {Nov},
  publisher = {American Physical Society},
  doi = {10.1103/PhysRevD.86.104018},
  url = {https://link.aps.org/doi/10.1103/PhysRevD.86.104018}
}

@article{Covariance1,
  title = {Covariance in models of loop quantum gravity: Spherical symmetry},
  author = {Bojowald, Martin and Brahma, Suddhasattwa and Reyes, Juan D.},
  journal = {Phys. Rev. D},
  volume = {92},
  issue = {4},
  pages = {045043},
  numpages = {16},
  year = {2015},
  month = {Aug},
  publisher = {American Physical Society},
  doi = {10.1103/PhysRevD.92.045043},
  url = {https://link.aps.org/doi/10.1103/PhysRevD.92.045043}
}

@article{Deformed2,
  title = {Deformed covariance in spherically symmetric vacuum models of loop quantum gravity: Consistency in Euclidean and self-dual gravity},
  author = {Bojowald, Martin and Brahma, Suddhasattwa and Ding, Ding and Ronco, Michele},
  journal = {Phys. Rev. D},
  volume = {101},
  issue = {2},
  pages = {026001},
  numpages = {22},
  year = {2020},
  month = {Jan},
  publisher = {American Physical Society},
  doi = {10.1103/PhysRevD.101.026001},
  url = {https://link.aps.org/doi/10.1103/PhysRevD.101.026001}
}

@article{midi,
doi = {10.1023/A:1026650212053},
url = {https://doi.org/10.1023/A:1026650212053},
year = {1999},
issn = {1572-9575},
month = {apr},
title = {Midisuperspace Models of Canonical Quantum Gravity},
journal = {International Journal of Theoretical Physics},
volume = {38},
issue = {4},
page = {1081},
author = {Torre, C. G.}
}

@article{Thiemann,
title = {Canonical quantization of spherically symmetric gravity in Ashtekar's self-dual representation},
journal = {Nuclear Physics B},
volume = {399},
number = {1},
pages = {211-258},
year = {1993},
issn = {0550-3213},
doi = {https://doi.org/10.1016/0550-3213(93)90623-W},
url = {https://www.sciencedirect.com/science/article/pii/055032139390623W},
author = {T. Thiemann and H.A. Kastrup}
}

@article{Kuchar,
  title = {Geometrodynamics of Schwarzschild black holes},
  author = {Kuchar, Karel V.},
  journal = {Phys. Rev. D},
  volume = {50},
  issue = {6},
  pages = {3961--3981},
  numpages = {0},
  year = {1994},
  month = {Sep},
  publisher = {American Physical Society},
  doi = {10.1103/PhysRevD.50.3961},
  url = {https://link.aps.org/doi/10.1103/PhysRevD.50.3961}
}

@article{Kastrup,
doi = {10.1088/0264-9381/17/15/311},
url = {https://dx.doi.org/10.1088/0264-9381/17/15/311},
year = {2000},
month = {aug},
publisher = {},
volume = {17},
number = {15},
pages = {3009},
author = {M Bojowald and  H A Kastrup},
title = {Symmetry reduction for quantized
diffeomorphism-invariant theories of connections},
journal = {Classical and Quantum Gravity}
}

@article{asier,
  title = {Nonsingular spherically symmetric black-hole model with holonomy corrections},
  author = {Alonso-Bardaji, Asier and Brizuela, David and Vera, Ra\"ul},
  journal = {Phys. Rev. D},
  volume = {106},
  issue = {2},
  pages = {024035},
  numpages = {26},
  year = {2022},
  month = {Jul},
  publisher = {American Physical Society},
  doi = {10.1103/PhysRevD.106.024035},
  url = {https://link.aps.org/doi/10.1103/PhysRevD.106.024035}
}

@article{zhang,
  title = {Black holes and covariance in effective quantum gravity},
  author = {Zhang, Cong and Lewandowski, Jerzy and Ma, Yongge and Yang, Jinsong},
  journal = {Phys. Rev. D},
  volume = {111},
  issue = {8},
  pages = {L081504},
  numpages = {7},
  year = {2025},
  month = {Apr},
  publisher = {American Physical Society},
  doi = {10.1103/PhysRevD.111.L081504},
  url = {https://link.aps.org/doi/10.1103/PhysRevD.111.L081504}
}

@book{wald,
    title = {General Relativity},
    author = {Robert M. Wald},
    isbn = {0-226-87033-2},
    year = {1984},
    publisher = {The University of Chicago Press},
}

@book{Thiemann_2007, place={Cambridge}, series={Cambridge Monographs on Mathematical Physics}, title={Modern Canonical Quantum General Relativity}, publisher={Cambridge University Press}, author={Thiemann, Thomas}, year={2007}, collection={Cambridge Monographs on Mathematical Physics}}

@book{Kiefer,
title={Quantum Gravity},
author={Kiefer, Claus},
year={2012},
publisher={Oxford University Press},
place={Oxford},
isbn={978-0-19-958520-5}
}

@article{Bojowald2004,
doi = {10.1088/0264-9381/21/15/008},
url = {https://dx.doi.org/10.1088/0264-9381/21/15/008},
year = {2004},
month = {jul},
publisher = {},
volume = {21},
number = {15},
pages = {3733},
author = {Martin Bojowald},
title = {Spherically symmetric quantum geometry: states and basic operators},
journal = {Classical and Quantum Gravity}
}

@article{Bojowald2006,
doi = {10.1088/0264-9381/23/6/015},
url = {https://dx.doi.org/10.1088/0264-9381/23/6/015},
year = {2006},
month = {mar},
publisher = {},
volume = {23},
number = {6},
pages = {2129},
author = {Martin Bojowald and Rafal Swiderski},
title = {Spherically symmetric quantum geometry: Hamiltonian constraint},
journal = {Classical and Quantum Gravity}
}

@article{Campiglia2007,
doi = {10.1088/0264-9381/24/14/007},
url = {https://dx.doi.org/10.1088/0264-9381/24/14/007},
year = {2007},
month = {jul},
publisher = {},
volume = {24},
number = {14},
pages = {3649},
author = {Miguel Campiglia and Rodolfo Gambini and Jorge Pullin},
title = {Loop quantization of spherically symmetric midi-superspaces},
journal = {Classical and Quantum Gravity}
}

@article{Tibrewala_2012,
doi = {10.1088/0264-9381/29/23/235012},
url = {https://dx.doi.org/10.1088/0264-9381/29/23/235012},
year = {2012},
month = {oct},
publisher = {IOP Publishing},
volume = {29},
number = {23},
pages = {235012},
author = {Rakesh Tibrewala},
title = {Spherically symmetric Einstein Maxwell theory and loop quantum gravity corrections},
journal = {Classical and Quantum Gravity}
}

@book{Kleinert,
author = {Kleinert, Hagen},
title = {Path Integrals in Quantum Mechanics, Statistics, Polymer Physics, and Financial Markets},
publisher = {WORLD SCIENTIFIC},
year = {2009},
doi = {10.1142/7305},
address = {},
edition   = {5th},
URL = {https://www.worldscientific.com/doi/abs/10.1142/7305},
eprint = {https://www.worldscientific.com/doi/pdf/10.1142/7305}
}

@article{LTB,
  title = {Lemaitre-Tolman-Bondi collapse from the perspective of loop quantum gravity},
  author = {Bojowald, Martin and Harada, Tomohiro and Tibrewala, Rakesh},
  journal = {Phys. Rev. D},
  volume = {78},
  issue = {6},
  pages = {064057},
  numpages = {30},
  year = {2008},
  month = {Sep},
  publisher = {American Physical Society},
  doi = {10.1103/PhysRevD.78.064057},
  url = {https://link.aps.org/doi/10.1103/PhysRevD.78.064057}
}

@article{DelAguila_2025,
doi = {10.1088/1361-6382/add079},
url = {https://doi.org/10.1088/1361-6382/add079},
year = {2025},
month = {may},
publisher = {IOP Publishing},
volume = {42},
number = {10},
pages = {105002},
author = {Del Águila, Juan Carlos and Morales-Técotl, Hugo A},
title = {Testing general covariance in effective models motivated by loop quantum gravity},
journal = {Classical and Quantum Gravity}
}

@article{brahma,
  title = {Effective line elements and black-hole models in canonical loop quantum gravity},
  author = {Bojowald, Martin and Brahma, Suddhasattwa and Yeom, Dong-han},
  journal = {Phys. Rev. D},
  volume = {98},
  issue = {4},
  pages = {046015},
  numpages = {16},
  year = {2018},
  month = {Aug},
  publisher = {American Physical Society},
  doi = {10.1103/PhysRevD.98.046015},
  url = {https://link.aps.org/doi/10.1103/PhysRevD.98.046015}
}

@misc{belfaqih,
      title={Black holes in effective loop quantum gravity: Covariant holonomy modifications}, 
      author={Idrus Husin Belfaqih and Martin Bojowald and Suddhasattwa Brahma and Erick I. Duque},
      year={2024},
      eprint={2407.12087},
      archivePrefix={arXiv},
      primaryClass={gr-qc},
      url={https://arxiv.org/abs/2407.12087}, 
}

@article{penrose,
  title = {Gravitational Collapse and Space-Time Singularities},
  author = {Penrose, Roger},
  journal = {Phys. Rev. Lett.},
  volume = {14},
  issue = {3},
  pages = {57--59},
  numpages = {0},
  year = {1965},
  month = {1},
  publisher = {American Physical Society},
  doi = {10.1103/PhysRevLett.14.57},
  url = {https://link.aps.org/doi/10.1103/PhysRevLett.14.57}
}

\end{document}